\documentclass[reprint, twocolumn, amsmath, amssymb, aps, prx, superscriptaddress]{revtex4-2}

\usepackage{graphicx}
\usepackage{dcolumn}
\usepackage{bm}
\usepackage{mathtools}
\usepackage{amsthm}
\usepackage{physics} 
\usepackage{xcolor}  
\usepackage{tabularx}
\usepackage{booktabs}

\usepackage[colorlinks]{hyperref}
\hypersetup{linkcolor=magenta,urlcolor=blue,citecolor=blue,pdfstartview={FitH},hyperfootnotes=false,unicode=true}

\def\be{\begin{equation}}
\def\ee{\end{equation}}
\def\bea{\begin{eqnarray}}
\def\eea{\end{eqnarray}}

\newcommand{\rhomem}{\rho^{\rm m}_{s_n}}    
\newcommand{\rhopred}{\rho^{\rm p}_{s_{n+1}}} 
\newcommand{\chimem}{\chi^{\rm m}_{t_{n+1}}}
\newcommand{\chipred}{\chi^{\rm p}_{t_{n+1}}}
\newcommand{\chidiss}{\chi^{\rm d}_{t_{n+1}}}

\begin{document}

\title{Thermodynamics of Quantum Reservoir Computing}

\author{Lixiang Ding}
\affiliation{School of Physics Science and Engineering, Tongji University, Shanghai 200092, China}

\author{Xingze Qiu}
\email{xingze@tongji.edu.cn}
\affiliation{School of Physics Science and Engineering, Tongji University, Shanghai 200092, China}

\date{\today}

\begin{abstract}

Quantum reservoir computing provides a framework for processing complex temporal data, yet its fundamental computational and energetic limits remain unresolved. Here, we establish a non-equilibrium thermodynamic framework that links the macroscopic predictive performance of driven open quantum systems to their microscopic energetic costs. By mapping Holevo capacities onto the Bogoliubov-Kubo-Mori geometric manifold, we analytically prove that the computational peak within the quantum critical region originates from a spectral resonance: the closing of the intrinsic energy gap forces the reservoir's internal transition frequencies to align with the chaotic drive. To evaluate the associated thermodynamic costs, we introduce quantum informational dissipation to quantify the non-predictive historical data retained by the reservoir. This allows us to derive a generalized Landauer bound for continuous temporal processing, which reveals a fundamental thermodynamic trade-off: the critical resonance that maximizes predictive capacity simultaneously maximizes informational dissipation and the irreversible work required for environmental erasure. Furthermore, coherence decomposition demonstrates that quantum coherences amplify predictive capacity without demanding additional mechanical work. These findings establish the fundamental energetic limits of quantum learning devices, providing theoretical principles for designing energy-efficient quantum neuromorphic hardware. 

\end{abstract}

\maketitle

\section{Introduction}
\label{sec:intro}

Processing complex temporal data and predicting chaotic dynamics are central challenges in machine learning and complex systems science \cite{Cheng_2015, Schotz_2026}. To circumvent the training bottlenecks of recurrent neural networks, reservoir computing utilizes the intrinsic dynamics of a physical system to project sequential inputs into a high-dimensional feature space \cite{Jaeger_2001, Maass_2002, Pathak_2018_PRL, Nakajima_2021_Book, Yan_2024_NC, Zolfagharinejad_2025_Nature}. Quantum reservoir computing (QRC) extends this paradigm by leveraging the exponential scaling of the Hilbert space, many-body entanglement, and open quantum dynamics to achieve processing capacities beyond classical limits \cite{Nakajima_2017, Ghosh_2019_PRL, Ghosh_2019_npjQI, Mujal_2021, Zambrini_2021_PRL, Angelatos_2021_PRX, Bravo_2022_PRXQuantum, Kubota_2023_PRR, Dudas_2023_npjQI, Senanian_2024_NC, Hu_2024_NC, Ahmed_2024_PRR, Kobayashi_2024_PRXQuantum, Kora_2024_PRA, Sannia_2024, Zhu_2025_PRR, Gotting_2025_PRL, Kobayashi_2025_NC, Danilo_2025_SciAdv, Kobayashi_2026_PRL, Hou_2026_PRL, Liu_2026_PRA, Bartolo_2026_npjQI, Paparelle_2026_NatPhoton, Sannia_2026_npjQI}. 

Despite successes in tasks such as quantum state preparation \cite{Ghosh_2019_PRL}, quantum tomography \cite{Ghosh_2019_npjQI, Angelatos_2021_PRX}, chaotic forecasting \cite{Kobayashi_2024_PRXQuantum, Ahmed_2024_PRR}, and entanglement witnessing \cite{Danilo_2025_SciAdv}, current frameworks primarily focus on the macroscopic capabilities of QRC. The fundamental thermodynamic mechanisms and microscopic physical origins underlying quantum temporal processing remain largely unresolved. Processing continuous data sequences requires dynamic memory updates, a process bounded by Landauer's principle of information erasure \cite{Landauer_1961, Sagawa_2009_PRL, Still_2012_PRL, Goold_2015_PRL, Van_2022_PRL, Hsieh_2025_PRL}. However, within non-equilibrium quantum dynamics, a rigorous connection between macroscopic predictive performance and environmental energy dissipation is absent. This gap is particularly evident regarding the computational peak observed in complex reservoirs. Empirical evidence indicates that optimal predictive performance occurs within the quantum critical region---a phenomenon traditionally referred to as computation at the ``edge of chaos'' \cite{Zambrini_2021_PRL, Kobayashi_2026_PRL}. Yet, an analytical microscopic framework explaining this performance maximization and its associated energetic constraints is incomplete. As QRC transitions toward diverse physical hardware implementations---including superconducting quantum processors \cite{Dudas_2023_npjQI, Senanian_2024_NC}, nuclear magnetic resonance systems \cite{Hou_2026_PRL}, Rydberg atom arrays \cite{Bravo_2022_PRXQuantum, Liu_2026_PRA}, and photonic integrated circuits \cite{Bartolo_2026_npjQI, Paparelle_2026_NatPhoton}---establishing a first-principles thermodynamic framework to resolve these basic physical questions is imperative. 

Here, we establish a non-equilibrium thermodynamic framework to resolve the physical origins and energetic constraints of QRC. By investigating an interacting many-body reservoir subjected to a continuous temporal sequence, we utilize the Holevo quantity \cite{Holevo_1973, Schumacher_1997_PRA, Holevo_1998, Holevo_2001_Book, Vedral_2002_RMP, Nielsen_Chuang_2010, Wilde_2017} to connect microscopic stochastic quantum trajectories with macroscopic predictive performance. By mapping these capacities onto the Bogoliubov-Kubo-Mori (BKM) geometric manifold \cite{Petz_1994, Petz_1996, Rudoy_2009}, we analytically prove that the computational peak within the quantum critical region originates from a spectral resonance: the closing of the intrinsic energy gap forces the internal transition frequencies of the reservoir to match the characteristic frequency of the chaotic drive. To evaluate the fundamental energetic costs of this optimal processing, we introduce quantum informational dissipation (QID), a metric that quantifies the non-predictive historical data retained by the quantum state. Using this framework, we derive a generalized Landauer bound, demonstrating that the non-equilibrium irreversible work produced during sequence processing is lower-bounded by the cumulative QID. This reveals a fundamental thermodynamic trade-off: the critical resonance that maximizes predictive capacity simultaneously maximizes informational dissipation and the irreversible work required for environmental erasure. Furthermore, we isolate the operational advantage of quantum coherence \cite{Baumgratz_2014_PRL, Streltsov_2017_RMP, Lecamwasam_2024_PRXQuantum}. By establishing a thermodynamically fair baseline across fully coherent and classically dephased protocols, we demonstrate that quantum coherences amplify the predictive capacity without demanding additional mechanical work. Validated across distinct paradigms of many-body quantum reservoirs, these results establish the fundamental energetic limits of quantum learning devices, providing theoretical principles for designing energy-efficient quantum neuromorphic hardware. 

The remainder of this paper is organized as follows. Section~\ref{sec:qrc} introduces the physical model and the Hamiltonian encoding protocol for QRC. Section~\ref{sec:info_capacities} defines the microscopic information capacities based on stochastic quantum trajectories. Section~\ref{sec:coherence} details the coherence decomposition of these informational metrics. In Sec.~\ref{sec:landauer}, we derive the generalized Landauer bound connecting macroscopic predictive performance to microscopic thermodynamic costs. Section~\ref{sec:quantum_advantage} isolates the quantum advantage provided by coherences. Section~\ref{sec:critical_thermo} analytically evaluates and numerically validates the computational peak within quantum critical regions. Finally, Sec.~\ref{sec:conclusions} concludes the paper with a discussion of the results and future perspectives.

\begin{figure}
	\centering
	\includegraphics[width=0.48\textwidth]{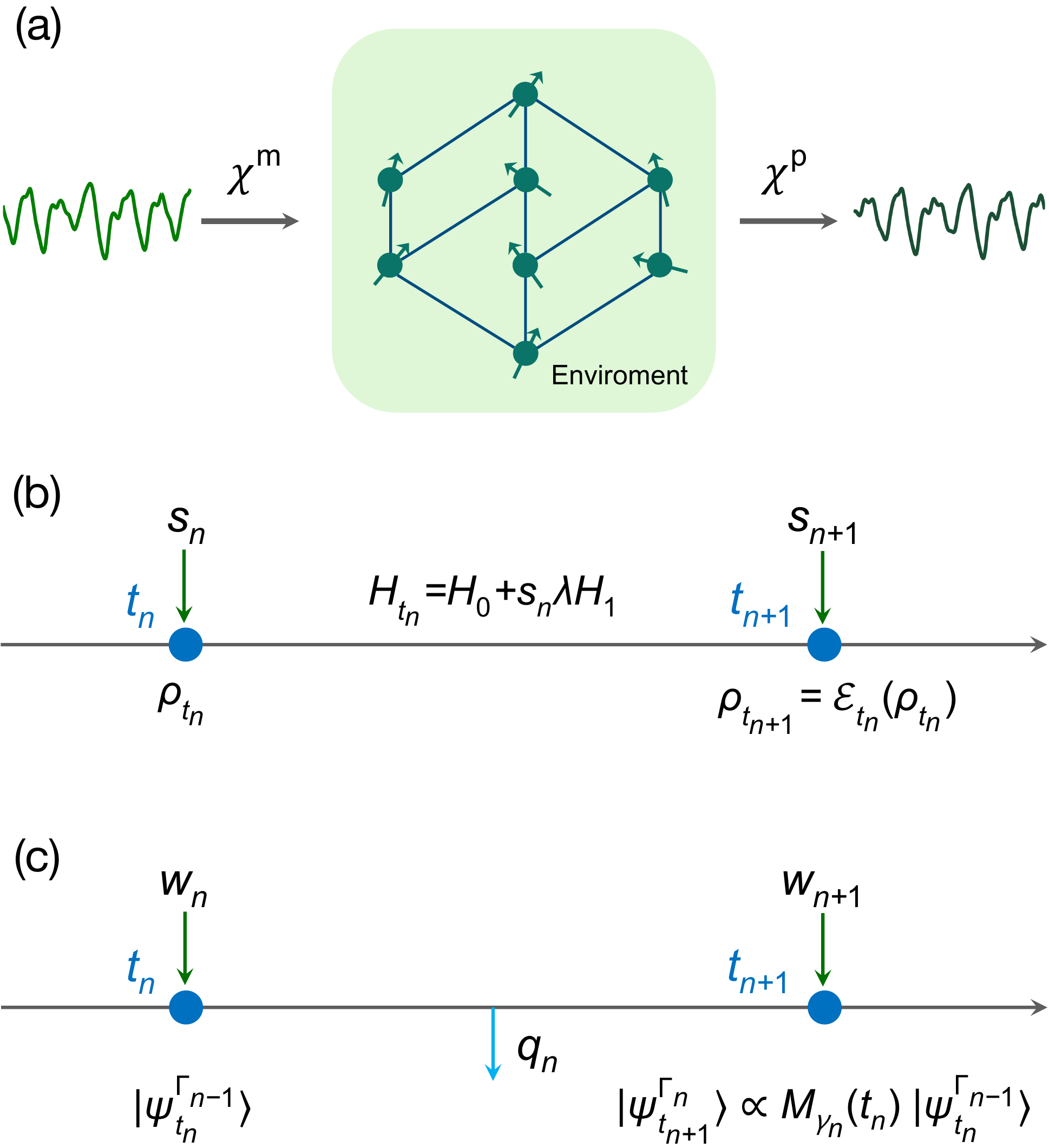}
	\caption{
	{\bf Thermodynamic framework of quantum reservoir computing.} 
	(a) Schematic of the continuous temporal processing paradigm. The input sequence is injected into a quantum many-body system interacting with an environment. The encoded historical information is quantified by the quantum memory Holevo capacity $\chi^{\rm m}$, while the correlation for predicting future targets is quantified by the predictive capacity $\chi^{\rm p}$. 
	(b) Hamiltonian encoding protocol. At each time step $t_n$, the input $s_n$ is injected via a sudden external field quench, updating the system Hamiltonian to $H_{t_n} = H_0 + s_n \lambda H_1$. The subsequent open-system dynamics evolve the macroscopic state from $\rho_{t_n}$ to $\rho_{t_{n+1}}$ via a completely positive and trace-preserving map $\mathcal{E}_{t_n}$. 
	(c) Microscopic stochastic trajectories. To evaluate the thermodynamic limits, the macroscopic map $\mathcal{E}_{t_n}$ is unraveled into an ensemble of Kraus operators $\{M_{\gamma_n}(t_n)\}$. This decomposes the open-system dynamics into individual quantum jump trajectories, where the microscopic pure state updates via $|\psi_{t_{n+1}}^{\Gamma_n}\rangle \propto M_{\gamma_n}(t_n) |\psi_{t_{n}}^{\Gamma_{n-1}}\rangle$, and $\Gamma_n = (\gamma_0, \gamma_1, \dots, \gamma_n)$ records the jump sequence. Tracking the microscopic stochastic work ($w_n$) and heat ($q_n$) fluxes links the non-equilibrium energetic costs to the macroscopic informational capacities ($\chi^{\rm m}$ and $\chi^{\rm p}$), yielding the generalized Landauer bound.
	}
	\label{Fig_1}
\end{figure}

\section{Quantum reservoir computing}
\label{sec:qrc}

We model the quantum reservoir as a driven open many-body system (Fig.~\ref{Fig_1}), driven by a discrete-time input sequence $\{s_0, s_1, \dots, s_n, \dots\}$. Although inputs are often injected via discrete qubit resetting \cite{Nakajima_2017}, such erase-and-write maps introduce artificial dissipation \cite{Gotting_2025_PRL}. To isolate the fundamental thermodynamic bounds of the information processing, we employ a Hamiltonian encoding protocol \cite{Bravo_2022_PRXQuantum, Kubota_2023_PRR, Sannia_2024, Zhu_2025_PRR, Gotting_2025_PRL}, advantageous for its continuous-time processing and direct experimental feasibility.
Under this protocol, each input $s_n$ couples to the system at time $t_n \equiv n\delta t$ via a sudden external field quench. This shifts the system Hamiltonian to $H_{t_n} = H_0 + s_n \lambda H_1$, where $H_0$ governs the intrinsic many-body dynamics, $H_1$ is the driving operator, and $\lambda$ is the coupling strength. During the subsequent interval $\delta t$, the continuous-time reservoir evolution follows a Markovian quantum master equation \cite{Lindblad_1976, Gorini_1976, Breuer_2007}:
\begin{equation}\label{eq:master_equation}
	\frac{\dd\rho}{\dd t} = \mathcal{L}_{t_n}(\rho) = -i \left[ H_{t_n}, \rho \right] + \mathcal{D}_{t_n}(\rho),
\end{equation}
where $\hbar = k_{\rm B} = 1$. Because the Hamiltonian remains constant during $\delta t$, integrating this time-independent Liouvillian defines the completely positive and trace-preserving (CPTP) map $\mathcal{E}_{t_n} = \exp(\mathcal{L}_{t_n} \delta t)$. This map dictates the macroscopic discrete-time state update $\rho_{t_{n+1}} = \mathcal{E}_{t_n}(\rho_{t_n})$ [Fig.~\ref{Fig_1}(b)]. 
While this map accommodates generic dissipators $\mathcal{D}_{t_n}(\rho)$, evaluating thermodynamic limits requires strict thermodynamic consistency \cite{Alicki_1976, Levy_2014, Hofer_2017, Cattaneo_2019, Dann_2021}. To correctly capture energy and entropy exchanges, the dissipator must be microscopically derived from the underlying system-bath interactions (Appendix~\ref{sec:appendix_discrete_thermalization_map}).

To extract computational utility, the reservoir projects the sequential inputs into a high-dimensional feature space through three stages: washout, training, and testing, spanning $N_{\text{wash}}$, $N_{\text{train}}$, and $N_{\text{test}}$ time steps, respectively \cite{Nakajima_2017}. During the washout phase, the reservoir is driven by a preliminary sequence without measurement. Environmental dissipation endows the reservoir with the fading memory property essential for processing streaming data \cite{Sannia_2024, Zhu_2025_PRR, Gotting_2025_PRL}. 
Following initialization, the training and testing stages extract macroscopic features at each time step $t_n$ by evaluating the expectation values of selected physical observables. 
A computational task targets an output sequence $\{\hat{y}_n\}$ corresponding to the inputs $\{s_n\}$. During training, the recorded expectation values are concatenated over $N_{\text{train}}$ steps to construct the training matrix $\mathbf{X}_{\text{train}}$, with corresponding targets forming the vector $\mathbf{\hat{y}}_{\text{train}}$. In QRC, optimization is restricted exclusively to a classical linear readout layer; the internal quantum dynamics remain fixed. To prevent overfitting, the readout weights $\mathbf{w}_{\text{opt}}$ are optimized via ridge regression, minimizing the error between the output $\mathbf{y}_{\text{train}} = \mathbf{X}_{\text{train}} \mathbf{w}_{\text{opt}}$ and the target $\mathbf{\hat{y}}_{\text{train}}$:
\begin{equation}\label{eq:ridge_regression}
\mathbf{w}_{\text{opt}} = \left( \mathbf{X}^T_{\text{train}} \mathbf{X}_{\text{train}} + \eta I \right)^{-1} \mathbf{X}^T_{\text{train}} \mathbf{\hat{y}}_{\text{train}},
\end{equation}
where $\eta$ is the regularization hyperparameter and $I$ is the identity matrix. During testing, the extracted expectation values form the testing matrix $\mathbf{X}_{\text{test}}$. The optimized weights generate the predicted outputs $\mathbf{y}_{\text{test}} = \mathbf{X}_{\text{test}} \mathbf{w}_{\text{opt}}$. The reservoir's computational capability is determined by the deviation between the prediction $\mathbf{y}_{\text{test}}$ and the target $\mathbf{\hat{y}}_{\text{test}}$, benchmarked using the normalized mean-squared error (NMSE):
\begin{equation}\label{eq:nmse}
	\text{NMSE} = \frac{\| \mathbf{y}_{\text{test}} - \mathbf{\hat{y}}_{\text{test}} \|^2}{\| \mathbf{\hat{y}}_{\text{test}} \|^2},
\end{equation}
where $\| \cdot \|$ is the Euclidean norm. A lower NMSE indicates higher predictive accuracy, serving as the standard metric for computational performance. 

While this macroscopic framework evaluates computational accuracy, it treats the quantum reservoir as an abstract algorithmic mapping, obscuring the underlying physical mechanisms. Dynamically updating a quantum memory to process continuous temporal sequences inevitably incurs non-equilibrium thermodynamic costs, including stochastic work, heat dissipation, and information processing. To establish the fundamental energetic limits of QRC, we must connect this macroscopic predictive performance to the microscopic stochastic thermodynamics.

\section{Information capacities}
\label{sec:info_capacities}

To link macroscopic predictive performance with microscopic thermodynamics, we unravel the CPTP map $\mathcal{E}_{t_n}$ into an ensemble of stochastic quantum trajectories [Fig.~\ref{Fig_1}(c)]. This unraveling corresponds to generalized quantum measurements continuously performed by the environment \cite{Brun_2000_PRA, Elouard_2017_npjQI}. The resulting measurement-induced wave-function collapses are the source of stochasticity and microscopic irreversibility. This process is formalized by defining Kraus operators $\{M_m(t_n)\}$ satisfying $\sum_m M_m^\dagger(t_n) M_m(t_n) = I$, which reconstruct the macroscopic map via $\mathcal{E}_{t_n}(\rho_{t_n}) = \sum_m M_m(t_n) \rho_{t_n} M_m^\dagger(t_n)$. Let $S_n = (s_0, s_1, \dots, s_n)$ denote a driving history drawn from the probability distribution $P(S_n)$, and $\Gamma_n = (\gamma_1, \gamma_2, \dots, \gamma_n)$ record the sequence of stochastic quantum jumps, where $\gamma_n$ indexes the realized Kraus operator $M_{\gamma_n}(t_n)$ at time $t_n$. The microscopic pure state $|\psi_{t_{n+1}}^{\Gamma_n}\rangle$ evolves with a jump probability $P(\gamma_n \mid S_n, \Gamma_{n-1}) = \| M_{\gamma_n}(t_n) |\psi_{t_{n}}^{\Gamma_{n-1}}\rangle \|^2$. Connecting these microscopic trajectories to macroscopic information processing requires statistical aggregation. The unconditional average state is defined by tracing over all driving histories and jump records:
\begin{equation}\label{eq:unconditional_state}
	\bar{\rho}_{t_{n+1}} = \sum_{S_n, \Gamma_n} P(S_n, \Gamma_n) \ket{\psi_{t_{n+1}}^{\Gamma_n}} \! \bra{\psi_{t_{n+1}}^{\Gamma_n}}.
\end{equation}

To quantify the reservoir's capacity to encode historical inputs and predict future targets, we define two conditional operational states. The memory state $\rhomem$ characterizes the reservoir's configuration conditioned on the present input $s_n$, constructed by statistically aggregating the microscopic trajectories over the past history $S_{n-1}$ and jump records $\Gamma_n$:
\begin{equation}\label{eq:memory_state}
	\rhomem = \sum_{S_{n-1}, \Gamma_n} P\left( S_{n-1}, \Gamma_n \mid s_n\right) \ket{\psi_{t_{n+1}}^{\Gamma_n}} \! \bra{\psi_{t_{n+1}}^{\Gamma_n}}.
\end{equation}
Similarly, predicting future dynamics relies on correlations between the reservoir's configuration and the forthcoming target $s_{n+1}$. The predictive state $\rhopred$ aggregates microscopic states conditioned exclusively on this target:
\begin{equation}\label{eq:predictive_state}
	\rhopred = \sum_{S_n, \Gamma_n} P\left(S_n, \Gamma_n \mid s_{n+1}\right) \ket{\psi_{t_{n+1}}^{\Gamma_n}} \! \bra{\psi_{t_{n+1}}^{\Gamma_n}}.
\end{equation}
Probability marginalization guarantees that both conditional ensembles reconstruct the unconditional average state (Appendix~\ref{sec:appendix_microscopic_trajectories}):
\begin{equation}\label{eq:consistency_check}
	\sum_{s_n} P(s_n) \rhomem = \sum_{s_{n+1}} P(s_{n+1}) \rhopred = \bar{\rho}_{t_{n+1}}.
\end{equation}
This consistency ensures that memory retention and target prediction are evaluated against the same physical baseline ensemble. 

The Holevo quantity $\chi$ determines the upper limit of classical information extractable from a quantum system via any generalized measurement \cite{Holevo_1973, Schumacher_1997_PRA, Holevo_1998, Holevo_2001_Book, Vedral_2002_RMP}. For an ensemble of quantum states $\rho_x$ prepared with classical probabilities $P(x)$, it is defined as $\chi = S(\bar{\rho}) - \sum_x P(x) S(\rho_x)$, where $\bar{\rho} = \sum_x P(x) \rho_x$ is the unconditional average density matrix and $S(\rho) = -\Tr(\rho \ln \rho)$ is the von Neumann entropy \cite{Nielsen_Chuang_2010, Wilde_2017}. In QRC, computational utility relies on extracting classical features from the quantum state. Evaluating specific observables introduces measurement-dependent bias. By bounding the extractable information independent of the measurement protocol, the Holevo quantity provides a hardware-agnostic metric to evaluate statistical correlations in driven open quantum systems. 

Applying this metric to continuous temporal sequences quantifies the theoretical limits of the reservoir's memory and predictive capacity. As formalized in Eqs.~\eqref{eq:memory_state} and \eqref{eq:predictive_state}, the conditional states $\rhomem$ and $\rhopred$ are defined via probabilistic marginalization, matching the required ensemble structure $\sum_x P(x) \rho_x$. 
Substituting these into the Holevo formula establishes two informational capacities. The quantum memory capacity $\chimem$ measures the historical information retained by the reservoir regarding the current drive $s_n$:
\begin{equation}\label{eq:chi_memory}
	\chimem = S(\bar{\rho}_{t_{n+1}}) - \sum_{s_n} P(s_n) S(\rhomem).
\end{equation}
Similarly, the quantum predictive capacity $\chipred$ measures the statistical correlation between the reservoir's dynamical state and the upcoming target $s_{n+1}$:
\begin{equation}\label{eq:chi_predictive}
	\chipred = S(\bar{\rho}_{t_{n+1}}) - \sum_{s_{n+1}} P(s_{n+1}) S(\rhopred).
\end{equation}
To evaluate the alignment between retained historical information and predictive correlation, we introduce the QID as the difference between these capacities:
\begin{equation}\label{eq:informational_dissipation}
	\chidiss = \chimem - \chipred.
\end{equation}
This metric serves as a comparative benchmark to assess the efficiency of the reservoir's information scheduling. 

Furthermore, because both conditional states $\rhomem$ and $\rhopred$ marginalize to the unconditional average $\bar{\rho}_{t_{n+1}}$, their Holevo capacities can be recast as the expected quantum relative entropy [$S(\rho \,\|\, \sigma) = \Tr(\rho \ln \rho) - \Tr(\rho \ln \sigma)$]. For the memory capacity, we obtain $\chimem = \sum_{s_n} P(s_n) S( \rhomem \| \bar{\rho}_{t_{n+1}})$. 
In the weak-coupling linear response regime, these conditional states act as infinitesimal perturbations around the unperturbed average, expressed as $\rho = \bar{\rho} + \lambda \delta \rho$, where $\lambda \ll 1$ is the coupling parameter and $\delta \rho$ is the traceless state variation. This allows us to approximate the relative entropy using the BKM metric \cite{Petz_1994}: $S(\bar{\rho} + \lambda \delta \rho \| \bar{\rho}) \approx \frac{\lambda^2}{2} g_{\text{BKM}}^{\bar{\rho}}(\delta \rho, \delta \rho)$. 
Applying this mapping projects the abstract Holevo capacities onto quadratic distances on the quantum state manifold. As detailed in Appendix~\ref{sec:geometric_formulation}, this geometric formulation yields explicit analytical expressions for the capacities, revealing the microscopic mechanisms driving computational performance.

While the preceding formulation focuses on the immediate step from $s_n$ to $s_{n+1}$, this framework generalizes to arbitrary temporal scales. By conditioning the microscopic ensembles on historical inputs with an arbitrary delay $\tau$ or on future targets at a prediction horizon $h$, we establish the $\tau$-step delayed memory capacity and the $h$-step predictive capacity. This extension evaluates the reservoir's capability to retain long-term temporal dependencies and anticipate distant dynamics. The construction and empirical validation of these multi-step capacities are detailed in Appendix~\ref{sec:multi_step_capacities}.

\section{Coherence decomposition}
\label{sec:coherence}

To determine whether the information capacities originate from classical population dynamics (diagonal distributions) or quantum superpositions (off-diagonal interference), we decompose the Holevo metric using the resource theory of coherence \cite{Baumgratz_2014_PRL, Streltsov_2017_RMP}. 
For a chosen basis $\mathcal{B} = \{\ket{m}\}$, the relative entropy of coherence for a state $\rho$ is $C(\rho) = S[\Delta(\rho)] - S(\rho)$ \cite{Baumgratz_2014_PRL}. Here, the dephasing map $\Delta(\rho) = \sum_m \langle m| \rho |m\rangle |m\rangle\!\langle m|$ removes all off-diagonal elements, reducing the state to its classical probability distribution along the diagonal. 
Consequently, the von Neumann entropy of the dephased state, $S[\Delta(\rho)]$, equals the Shannon entropy of the corresponding classical distribution in basis $\mathcal{B}$. 

For a generic ensemble of quantum states $\rho_x$ prepared with classical probabilities $P(x)$, the Holevo quantity $\chi$ partitions into classical and quantum components \cite{Lecamwasam_2024_PRXQuantum} (Appendix~\ref{sec:appendix_coherence}):
\begin{equation}\label{eq:holevo_decomposition}
	\chi = \mathcal{I} + \mathcal{C}.
\end{equation}
The first component is the classical Shannon mutual information, $\mathcal{I} = S[\Delta(\bar{\rho})] - \sum_x P(x) S[\Delta(\rho_x)]$, which represents the information extractable exclusively via the diagonal populations. The second component is the ensemble coherence, $\mathcal{C} = \sum_x P(x) C(\rho_x) - C(\bar{\rho})$. It measures the excess encoding capacity provided by off-diagonal superpositions, an operational advantage inaccessible to classical reservoirs. 
Due to the convexity of the relative entropy of coherence \cite{Baumgratz_2014_PRL}, the ensemble coherence is non-negative ($\mathcal{C} \ge 0$). This establishes the informational hierarchy $\chi \ge \mathcal{I}$, demonstrating that dephasing discards temporal features encoded in off-diagonal quantum superpositions.

Applying this bipartition to our framework, we evaluate the classical and quantum components of both the memory and predictive ensembles. As established in Eq.~\eqref{eq:consistency_check}, both ensembles share the same unconditional average state $\bar{\rho}_{t_{n+1}}$. This shared statistical structure allows us to define the classical information capacities: $\mathcal{I}^{\rm m}_{t_{n+1}} = S[\Delta(\bar{\rho}_{t_{n+1}})] - \sum_{s_n} P(s_n) S[\Delta(\rhomem)]$ and $\mathcal{I}^{\rm p}_{t_{n+1}} = S[\Delta(\bar{\rho}_{t_{n+1}})] - \sum_{s_{n+1}} P(s_{n+1}) S[\Delta(\rhopred)]$. Similarly, the corresponding ensemble coherences are formulated as: $\mathcal{C}^{\rm m}_{t_{n+1}} = \sum_{s_n} P(s_n) C(\rhomem) - C(\bar{\rho}_{t_{n+1}})$ and $\mathcal{C}^{\rm p}_{t_{n+1}} = \sum_{s_{n+1}} P(s_{n+1}) C(\rhopred) - C(\bar{\rho}_{t_{n+1}})$. Using these constituent capacities, we define the classical population dissipation as $\mathcal{D}^{\rm c}_{t_{n+1}} = \mathcal{I}^{\rm m}_{t_{n+1}} - \mathcal{I}^{\rm p}_{t_{n+1}}$ and the quantum coherent dissipation as $\mathcal{D}^{\rm q}_{t_{n+1}} = \mathcal{C}^{\rm m}_{t_{n+1}} - \mathcal{C}^{\rm p}_{t_{n+1}}$. Consequently, the total quantum informational dissipation partitions into these two components:
\begin{equation}\label{eq:dissipation_decomposition}
	\chidiss = \mathcal{D}^{\rm c}_{t_{n+1}} + \mathcal{D}^{\rm q}_{t_{n+1}}.
\end{equation}
This decomposition separates the total informational dissipation into classical and quantum parts, establishing a foundation to evaluate the quantum advantage.

\section{Generalized Landauer bound}
\label{sec:landauer}

Landauer's principle establishes the thermodynamic link between irreversible information processing and environmental heat dissipation \cite{Landauer_1961}, serving as a cornerstone for evaluating computing efficiency across classical and quantum regimes \cite{Sagawa_2009_PRL, Still_2012_PRL, Goold_2015_PRL, Van_2022_PRL, Hsieh_2025_PRL}. 
In QRC, however, the continuous assimilation of complex temporal sequences creates a complex physical scenario: the system is persistently driven far from equilibrium while simultaneously managing the real-time retention and predictive scheduling of non-equilibrium data. To resolve the energetic limits under these dynamical conditions, the standard Landauer bound must be generalized to incorporate the macroscopic predictive utility of the physical device. 
We achieve this by applying stochastic thermodynamics directly to the unraveled microscopic quantum trajectories \cite{Elouard_2017_npjQI, Alonso_2016_PRL, Strasberg_2019_PRE}. Tracking the fluctuating thermodynamic work and heat at the single-trajectory level enables us to decompose the continuous processing cycle into two sequential stages: instantaneous information injection and subsequent thermal relaxation [Fig.~\ref{Fig_1}(c)]. 

During the information injection stage at $t_{n+1}$, the signal $s_{n+1}$ is encoded via a sudden external quench. The global Hamiltonian abruptly shifts from $H_{t_n}$ to $H_{t_{n+1}}$. As this quench is instantaneous, the microscopic pure state $|\psi_{t_{n+1}}^{\Gamma_n}\rangle$ remains momentarily frozen. Consequently, the microscopic stochastic work $w_{n+1}$ performed on the reservoir equals the energy differential of the driving field: $w_{n+1} = \langle\psi_{t_{n+1}}^{\Gamma_n}|H_{t_{n+1}}|\psi_{t_{n+1}}^{\Gamma_n}\rangle - \langle\psi_{t_{n+1}}^{\Gamma_n}|H_{t_n}|\psi_{t_{n+1}}^{\Gamma_n}\rangle$. Averaging over all classical driving sequences and quantum jump records yields the macroscopic work $W_{n+1} = \sum_{S_{n+1}, \Gamma_{n}} P(S_{n+1}, \Gamma_{n}) w_{n+1}$. To isolate the penalty incurred by this quench, we utilize the non-equilibrium free energy $\mathcal{F}(\rho) = \Tr(\rho H) - \beta^{-1} S(\rho)$ \cite{Gaveau_1997, Sivak_2012_PRL, Parrondo_2015_NP}. This functional relates to the equilibrium free energy $F^{\text{eq}}$ via $\mathcal{F}(\rho) = F^{\text{eq}} + \beta^{-1} S(\rho \,\|\, \rho^{\text{eq}})$, where $\rho^{\text{eq}}$ is the instantaneous thermal Gibbs state. We define the instantaneous average irreversible work, $W^{\text{irr}}_{n+1}$, as the deviation of the actual work from the macroscopic free energy difference: $W^{\text{irr}}_{n+1} = W_{n+1} - \Delta \mathcal{F}_{n+1}$ \cite{Parrondo_2015_NP}. Here, $\Delta \mathcal{F}_{n+1}$ represents the non-equilibrium free energy change between the predictive states immediately following the quench and the memory states immediately preceding it. Explicitly, $\Delta \mathcal{F}_{n+1} = \sum_{s_{n+1}} P(s_{n+1}) [{\rm Tr}(\rhopred H_{t_{n+1}}) - \beta^{-1} S(\rhopred )] - \sum_{s_n} P(s_n) [{\rm Tr}(\rhomem H_{t_n}) - \beta^{-1} S(\rhomem)]$. Subtracting this boundary difference from $W_{n+1}$ cancels the energetic expectation values, reducing the instantaneous irreversible work to an entropic penalty: $\beta W^{\text{irr}}_{n+1} = \sum_{s_{n+1}} P(s_{n+1}) S(\rhopred) - \sum_{s_n} P(s_n) S(\rhomem)$ (Appendix~\ref{sec:appendix_landauer}). Recalling Eq.~\eqref{eq:informational_dissipation}, this entropic cost translates directly to the QID:
\begin{equation}\label{eq:work_dissipation_equivalence}
	\beta W^{\text{irr}}_{n+1} = \chidiss.
\end{equation}
This establishes a fundamental identity: the average irreversible thermodynamic cost required to drive the memory states into the predictive states equals the QID.

During the subsequent relaxation stage, the reservoir thermalizes with the environment under the fixed Hamiltonian $H_{t_{n+1}}$ for a duration $\delta t$. Because the Hamiltonian remains constant, the microscopic quantum heat $q_{n+1}$ exchanged with the bath equals the change in the system's energy during this interval: $q_{n+1} = \langle\psi_{t_{n+2}}^{\Gamma_{n+1}}|H_{t_{n+1}}|\psi_{t_{n+2}}^{\Gamma_{n+1}}\rangle - \langle\psi_{t_{n+1}}^{\Gamma_n}|H_{t_{n+1}}|\psi_{t_{n+1}}^{\Gamma_n}\rangle$. By defining the change in the microscopic internal energy across the entire processing step as $\Delta u_{n+1} = \langle\psi_{t_{n+2}}^{\Gamma_{n+1}}|H_{t_{n+1}}|\psi_{t_{n+2}}^{\Gamma_{n+1}}\rangle - \langle\psi_{t_{n+1}}^{\Gamma_n}|H_{t_n}|\psi_{t_{n+1}}^{\Gamma_n}\rangle$, the energy exchange satisfies the first law of thermodynamics at the single-trajectory level: $\Delta u_{n+1} = w_{n+1} + q_{n+1}$. Averaging over the statistical ensemble yields the macroscopic internal energy change $\Delta U_{n+1} = \sum_{S_{n+1}, \Gamma_{n+1}} P(S_{n+1}, \Gamma_{n+1}) \Delta u_{n+1}$ and the macroscopic heat $Q_{n+1} = \sum_{S_{n+1}, \Gamma_{n+1}} P(S_{n+1}, \Gamma_{n+1}) q_{n+1}$, preserving the macroscopic first law: $\Delta U_{n+1} = W_{n+1} + Q_{n+1}$ (Appendix~\ref{sec:appendix_landauer}). 
While no external mechanical work is performed during this fixed-Hamiltonian interval, spontaneous thermalization towards the environment entails irreversible entropy production. The energetic equivalent of this dissipation is quantified by the relaxation average irreversible work $W^{\text{relax}}_{n+1}$, defined as the negative change in the macroscopic non-equilibrium free energy during $\delta t$. This evaluates to $W^{\text{relax}}_{n+1} = \beta^{-1} \sum_{s_{n+1}} P(s_{n+1}) [ S(\rhopred \,\|\, \rho^{\text{eq}}_{s_{n+1}}) - S(\rho^{\text{m}}_{s_{n+1}} \,\|\, \rho^{\text{eq}}_{s_{n+1}}) ]$, where $\rho^{\text{eq}}_{s_{n+1}} = {\rm exp}(-\beta H_{t_{n+1}})/{\rm Tr} [{\rm exp}(-\beta H_{t_{n+1}})]$. The thermal relaxation from $t_{n+1}$ to $t_{n+2}$ is governed by the CPTP map $\mathcal{E}_{t_{n+1}}$. This map dynamically evolves the conditional predictive state into the subsequent memory state, $\mathcal{E}_{t_{n+1}}(\rhopred) = \rho^{\text{m}}_{s_{n+1}}$, while leaving the instantaneous Gibbs state invariant, $\mathcal{E}_{t_{n+1}}(\rho^{\text{eq}}_{s_{n+1}}) = \rho^{\text{eq}}_{s_{n+1}}$. According to the monotonicity of quantum relative entropy under CPTP maps [$S(\mathcal{E}(\rho) \| \mathcal{E}(\sigma)) \le S(\rho \| \sigma)$] \cite{Wilde_2017}, the relative distance to the thermal equilibrium decreases. This guarantees that the relaxation dissipation is non-negative, $W^{\text{relax}}_{n+1} \ge 0$ (Appendix~\ref{sec:appendix_landauer}). 

To determine the global energetic limits over a complete computational task of duration $N\delta t$, we sum the irreversible work from both the injection and relaxation stages: $W^{\text{irr}}_{\text{tot}} = \sum_{n=0}^{N-1} ( W^{\text{irr}}_{n+1} + W^{\text{relax}}_{n+1})$. 
Discarding the non-negative relaxation contributions yields a macroscopic inequality bounded by the informational dissipation:
\begin{equation}\label{eq:total_work_bound}
	\beta W^{\text{irr}}_{\text{tot}} \ge \chi^{\rm d}_{\rm tot},
\end{equation}
where $\chi^{\rm d}_{\rm tot} = \sum_{n=0}^{N-1} \chidiss$ is the accumulated QID. 
To establish a direct link between the total irreversible work and the environmental heat dissipation, we define the global thermodynamic quantities integrated over the entire processing sequence: the total macroscopic work $W_{\text{tot}} = \sum_{n=0}^{N-1} W_{n+1}$, the total internal energy change $\Delta U_{\text{tot}} = \sum_{n=0}^{N-1} \Delta U_{n+1}$, and the total heat dissipated into the thermal bath $Q_{\text{diss}} = - \sum_{n=0}^{N-1} Q_{n+1}$. These quantities obey the macroscopic first law of thermodynamics: $W_{\text{tot}} = \Delta U_{\text{tot}} + Q_{\text{diss}}$. Next, we define the net change in the conditional von Neumann entropy of the reservoir between the initial and final states as $\Delta S_{\text{sys}} = \sum_{s_0} P(s_0) S(\rho^{\rm m}_{s_0}) - \sum_{s_N} P(s_N) S(\rho^{\rm m}_{s_{N}})$. Consequently, the total change in the macroscopic non-equilibrium free energy is constructed as $\Delta \mathcal{F}_{\text{tot}} = \Delta U_{\text{tot}} + \beta^{-1} \Delta S_{\text{sys}}$. 
By definition, the total irreversible work characterizes the excess work injected beyond this free energy difference: $W^{\text{irr}}_{\text{tot}} = W_{\text{tot}} - \Delta \mathcal{F}_{\text{tot}}$. We thus have $\beta W^{\text{irr}}_{\text{tot}} = \beta Q_{\text{diss}} - \Delta S_{\text{sys}}$. 
Substituting this relation into Eq.~\eqref{eq:total_work_bound} yields the primary thermodynamic bound: 
\begin{equation}\label{eq:quantum_landauer}
	\beta Q_{\text{diss}} \ge \Delta S_{\text{sys}} + \chi^{\rm d}_{\rm tot}.
\end{equation}
This inequality establishes a generalized Landauer bound for continuous quantum prediction. It demonstrates that Landauer's original quasi-static erasure bound ($\beta Q_{\text{diss}} \ge \Delta S_{\text{sys}}$) is insufficient for complex temporal processing: the total heat dissipated into the environment is augmented by the thermodynamic penalty of encoding non-predictive historical information.

\begin{figure*}
	\includegraphics[width=0.93\textwidth]{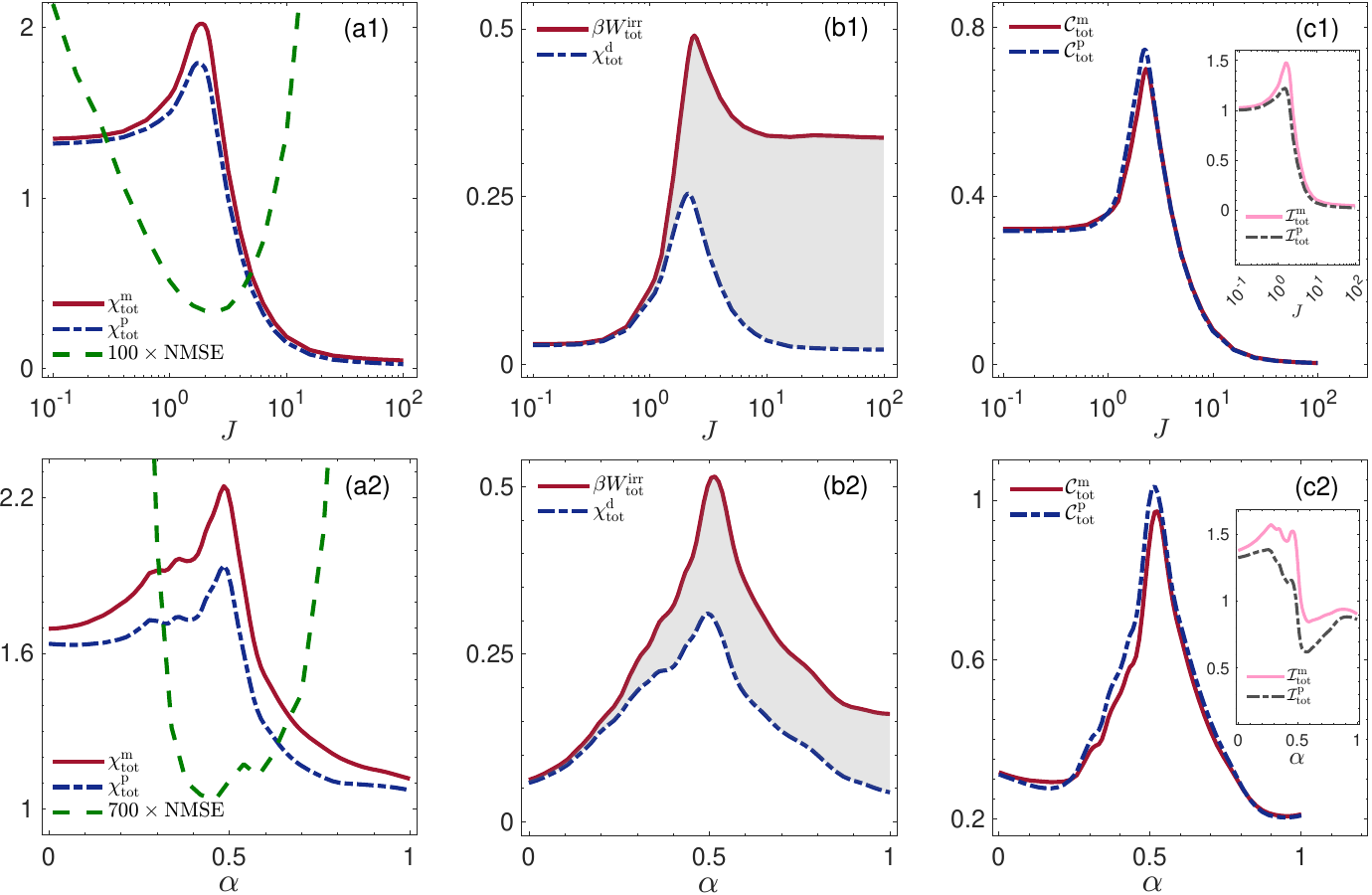}
\caption{
	{\bf Thermodynamic trade-offs and quantum advantage in quantum critical regions.} 
	The top row (a1--c1) presents results for the disordered transverse-field Ising model (TFIM), while the bottom row (a2--c2) corresponds to the augmented cluster model.
	(a) Macroscopic task performance. The Holevo capacities for memory ($\chi^{\rm m}_{\rm tot}$, red solid lines) and prediction ($\chi^{\rm p}_{\rm tot}$, blue dash-dotted lines) maximize within the quantum critical regions. This aligns with the global minimum of the forecasting error (NMSE, green dashed lines) evaluated via a full Pauli basis, connecting microscopic information capacities with macroscopic computational utility. 
	(b) Validation of the generalized Landauer bound. The cumulative average irreversible work $\beta W^{\text{irr}}_{\text{tot}}$ (red solid lines) is lower-bounded by the accumulated quantum informational dissipation $\chi^{\rm d}_{\rm tot}$ (blue dash-dotted lines). Both quantities exhibit peaks within their respective quantum critical regions ($J \approx 2.5$ and $\alpha \approx 0.5$). 
	(c) Coherence decomposition. The ensemble coherence of the memory ($\mathcal{C}^{\rm m}_{\rm tot}$, red solid lines) and predictive ($\mathcal{C}^{\rm p}_{\rm tot}$, blue dash-dotted lines) states peaks within the critical regions and satisfies $\mathcal{C}^{\rm p}_{\rm tot} > \mathcal{C}^{\rm m}_{\rm tot}$ across broad parameter regimes. Inset: The classical mutual information ($\mathcal{I}^{\rm m,p}_{\rm tot}$) lacks analogous critical scaling, demonstrating that the computational peak is driven by quantum coherence. 
	{\bf Simulation parameters}: System size $L = 6$, inverse temperature $\beta = 1$, thermal relaxation rate $\gamma_0 = 0.1$, temporal driving interval $\delta t = 1$, regularization hyperparameter $\eta=10^{-5}$, and longitudinal driving amplitude $\lambda = 0.05$. The NMSE is evaluated with $N_{\text{wash}} = 500$, $N_{\text{train}} = 2000$, and $N_{\text{test}} = 2000$, averaged over 500 sequences; for the disordered TFIM, each sequence employs a distinct random Hamiltonian realization. Thermodynamic and informational metrics are evaluated over $N_{\text{eval}} = 2000$ steps following an initial $N_{\text{wash}} = 500$ steps. Averages are taken over 5000 random sequences; for the disordered TFIM, this comprises 100 random Hamiltonian realizations with 5000 sequences each. Full numerical details are provided in Appendix~\ref{sec:numerical_methods}.
}
	\label{Fig_2}
\end{figure*}

Applying our coherence decomposition [Eq.~\eqref{eq:dissipation_decomposition}] to this generalized bound isolates the energetic consequences of quantum interference: $\beta Q_{\text{diss}} \ge \Delta S_{\text{sys}} + \left( \mathcal{D}^{\rm c}_{\rm tot} + \mathcal{D}^{\rm q}_{\rm tot} \right)$, 
where $\mathcal{D}^{\rm c}_{\rm tot} = \sum_{n=0}^{N-1} \mathcal{D}^{\rm c}_{t_{n+1}}$ and $\mathcal{D}^{\rm q}_{\rm tot} = \sum_{n=0}^{N-1} \mathcal{D}^{\rm q}_{t_{n+1}}$. 
This decomposition provides a basis to evaluate the thermodynamic consequences of quantum coherence. When the classical driving sequence is a Markov process, the coherent dissipation is non-negative ($\mathcal{D}^{\rm q}_{\rm tot} \ge 0$, proved in Appendix~\ref{sec:appendix_coherence}). In this scenario, unaligned quantum superpositions are irreversibly collapsed by environmental decoherence, adding a non-negative penalty to the thermodynamic threshold. Conversely, when processing complex, non-Markovian temporal sequences, the reservoir dynamics can yield $\mathcal{D}^{\rm q}_{\rm tot} < 0$, as demonstrated in the subsequent examples [Fig.~\ref{Fig_2}(c)]. 
In this regime, a negative $\mathcal{D}^{\rm q}_{\rm tot}$ lowers the theoretical minimum of the irreversible heat dissipation. By pulling the global dissipation threshold below the classical population limit ($\mathcal{D}^{\rm c}_{\rm tot}$), this reduction in the fundamental bound opens a permissible thermodynamic window for rendering quantum reservoirs energetically superior to their classical counterparts.

\section{Quantum advantage}
\label{sec:quantum_advantage}

While the thermodynamic benefit of negative coherent dissipation ($\mathcal{D}^{\rm q}_{\rm tot} < 0$) is sequence-dependent, we establish an unconditional quantum advantage: the ability of quantum coherences to amplify predictive performance without requiring additional mechanical work. To isolate this advantage, we contrast two paradigms: a fully coherent quantum readout and a classically dephased readout. Both paradigms undergo identical non-equilibrium open-system dynamics $\mathcal{E}_{t_n}$ prior to information extraction, ensuring a thermodynamically fair baseline for comparison. 

In the coherent paradigm, the informational capacities are evaluated using the full density matrix, yielding the Holevo capacities for memory ($\chimem$) and prediction ($\chipred$). Conversely, the classical paradigm evaluates these capacities without quantum interference. Immediately prior to readout, the system is subjected to a dephasing channel $\Delta$, projecting the state into the basis $\mathcal{B}$. The extractable capacities are thus restricted to the classical mutual information residing within the diagonal populations, denoted as $\mathcal{I}^{\rm m}_{t_{n+1}}$ and $\mathcal{I}^{\rm p}_{t_{n+1}}$. Because the ensemble coherence is non-negative ($\mathcal{C} \ge 0$), this terminal dephasing enforces the informational hierarchy: $\chimem \ge \mathcal{I}^{\rm m}_{t_{n+1}}$ and $\chipred \ge \mathcal{I}^{\rm p}_{t_{n+1}}$. 

We next determine whether achieving this enhanced predictive capacity requires additional mechanical work. During the signal injection step $s_n \to s_{n+1}$, the microscopic stochastic work performed on the reservoir is $w_{n+1} = \lambda(s_{n+1} - s_n) \langle\psi_{t_{n+1}}^{\Gamma_n}|H_1|\psi_{t_{n+1}}^{\Gamma_n}\rangle$. To establish thermodynamic equivalence, we align the driving operator $H_1$ to be diagonal in the dephasing basis $\mathcal{B}$. Because $H_1$ lacks off-diagonal elements, its expectation value is independent of quantum coherences, satisfying $\Tr[\rho H_1] = \Tr[\Delta(\rho) H_1]$ for the instantaneous pure density matrix $\rho = |\psi_{t_{n+1}}^{\Gamma_n}\rangle\! \langle\psi_{t_{n+1}}^{\Gamma_n}|$. Consequently, the instantaneous mechanical work depends solely on the classical populations. 
Because both paradigms evolve identically prior to the terminal readout, their pre-readout states possess identical diagonal populations $\Delta(\rho)$. They therefore incur the same microscopic work during each information injection. Integrating this over the complete temporal sequence yields an identical total macroscopic work $W_{\text{tot}}$ for both paradigms. Combining this energetic equivalence with the informational hierarchy demonstrates that mobilized quantum coherences amplify the predictive capacity ($\chipred \ge \mathcal{I}^{\rm p}_{t_{n+1}}$) without requiring additional mechanical work.

\section{Quantum critical thermodynamics}
\label{sec:critical_thermo}

To validate our theoretical framework, we evaluate the thermodynamic and computational performance of driven open many-body systems across distinct physical regimes. We consider two reservoir architectures. The first is a fully connected disordered transverse-field Ising model (TFIM) \cite{Zambrini_2021_PRL, Kora_2024_PRA, Sannia_2024}. Its intrinsic Hamiltonian is $H_0 = \sum_{i>j=1}^L J_{ij} \sigma_i^z \sigma_j^z + h \sum_{i=1}^L \sigma_i^x$, where the transverse magnetic field is $h = 1$, and the random long-range couplings $J_{ij}$ are sampled uniformly from the interval $[-J/2, J/2]$. Tuning the coupling strength $J$ drives a phase transition separating an ergodic phase from a many-body localized phase. The second architecture is a clean, one-dimensional cluster model augmented with a transverse magnetic field \cite{Wolf_2006_PRL, Smith_2022_PRR}, defined by $H_0 = -J_{zz} \sum_{i=1}^L \sigma_i^z \sigma_{i+1}^z - h_x \sum_{i=1}^L \sigma_i^x + J_{zxz} \sum_{i=1}^L \sigma_i^z \sigma_{i+1}^x \sigma_{i+2}^z$. Here, the nearest-neighbor Ising interaction is fixed at $J_{zz} = 0.1$, while the three-body cluster interaction and the transverse field are parameterized complementarily via a tuning parameter $\alpha \in [0, 1]$ as $J_{zxz} = (1 - J_{zz})\alpha$ and $h_x = (1 - J_{zz})(1 - \alpha)$. This interpolation drives the system across a symmetry-protected topological phase transition \cite{Smith_2022_PRR}.

For both architectures, the temporal data is generated by the chaotic Mackey-Glass (MG) equation \cite{MG_1977} and injected into the reservoir via a longitudinal driving field $H_1 = \sum_{i=1}^L \sigma_i^z$. The coupling strength is fixed at $\lambda = 0.05$ to ensure the drive acts as a physical perturbation. The environmental dissipation is governed by a discrete thermalization map that guarantees thermodynamic consistency (Appendix~\ref{sec:appendix_discrete_thermalization_map}). To evaluate the information flow and energetic costs, we process statistical ensembles of these chaotic sequences. For each sequence, an initial washout period of $N_{\text{wash}}$ steps is discarded to eliminate initialization transients. The microscopic memory and predictive capacities, along with their associated thermodynamic quantities, are accumulated over the subsequent $N_{\text{eval}}$ evaluation steps (Appendix~\ref{sec:numerical_methods}).

To determine the physical mechanisms driving the computational performance, we utilize the BKM metric to evaluate the informational capacities in the linear response limit. Taking the memory capacity as an example, the analytical expansion evaluates to (Appendix~\ref{subsec:microscopic_origin}):
\begin{align}
	\chi^{\rm m}_{t_{n+1}} &\approx \frac{\lambda^2}{2} \left[ \beta \sum_{j \neq k} G(\Delta E_{jk}) \frac{p_k - p_j}{E_j - E_k} F_{jk} \left| \!\mel{j}{H_1}{k} \right|^2 \right. \nonumber \\
	& \left. + \beta^2 P_{\text{th}}^2 G(0) \sum_j  p_j \left| \!\mel{j}{H_1 - \langle H_1 \rangle_0}{j} \right|^2 \right].
	\label{eq:analytical_chi_mem}
\end{align}
Here, $E_j$ and $p_j=\exp(-\beta E_j)/Z_0$ are the eigenenergies and thermal populations of the Hamiltonian $H_0$, $\Delta E_{jk} = E_j - E_k$ is the energy gap, $P_{\text{th}}$ is the thermalization probability per time step, $F_{jk} = |1 - (1-P_{\text{th}})\exp(-i \Delta E_{jk} \delta t)|^2$, and $\langle H_1 \rangle_0 = \sum_j p_j\!\mel{j}{H_1}{j}$ is the thermal expectation value. The explicit dependence on the inverse temperature $\beta$ indicates that the capacity increases as the system cools, rendering lower temperatures more effective for temporal processing. The dynamic temporal accumulation factor is given by $G(\Delta E_{jk}) = \sigma_{\rm s}^{-2} | \sum_{\tau=0}^\infty C_{\rm a}(\tau) (1-P_{\text{th}})^\tau \exp(-i \Delta E_{jk} \tau\delta t) |^2$, where $C_{\rm a}(\tau)$ is the auto-covariance function of the input sequence and $\sigma_{\rm s}^2$ is its variance. The predictive capacity $\chi^{\rm p}_{t_{n+1}}$ admits an analogous analytical expansion. 

The factor $G(\Delta E_{jk})$ functions as a frequency-selective filter. The chaotic MG sequence is characterized by a zero mean and a low-frequency dominant spectral peak at $\omega_{\rm s}$. Consequently, $G(\Delta E_{jk})$ imposes a resonance condition. For diagonal elements ($j=k$), the zero-frequency accumulation $G(0)$ is suppressed by destructive interference. For off-diagonal elements ($j \neq k$), $G(\Delta E_{jk})$ reaches a significant value only when the reservoir's internal energy gap matches the driving frequency ($\Delta E_{jk} \approx \omega_{\rm s}$), which eliminates phase oscillations and enables constructive temporal accumulation. 
Because the characteristic frequency $\omega_{\rm s}$ is small, this resonance condition is satisfied within the quantum critical region. Deep within a fully gapped phase, the intrinsic energy gaps are larger than the driving frequency ($\Delta E \gg \omega_{\rm s}$), resulting in a frequency mismatch that suppresses accumulation ($G \approx 0$). As the system is tuned toward a quantum phase transition, the primary energy gap closes. This collapse forces the internal transition frequencies to sweep through the low-frequency domain, aligning with $\omega_{\rm s}$ and triggering the spectral resonance. While this dynamical resonance acts as the foundational enabler, the total capacity is modulated by other physical factors: $(p_k - p_j)/(E_j - E_k)$ favoring small gaps, $F_{jk}$ behaving oppositely, and $| \!\mel{j}{H_1}{k} |^2$ increasing within the critical region. Detailed derivations and the synthesis of these mechanisms are provided in Appendix~\ref{subsec:microscopic_origin}. 

The numerical results confirm this theoretical mechanism and the corresponding thermodynamic constraints (Fig.~\ref{Fig_2}). Across both phase diagrams, the total memory ($\chi^{\rm m}_{\rm tot}=\sum_{n=0}^{N-1} \chimem$) and predictive ($\chi^{\rm p}_{\rm tot}=\sum_{n=0}^{N-1} \chipred$) capacities simultaneously reach their maxima within the quantum critical regions [Fig.~\ref{Fig_2}(a1, a2)]. This informational peak aligns with a minimum in the one-step-ahead forecasting error (NMSE); results for multi-step prediction horizons are detailed in Appendix~\ref{sec:multi_step_capacities}. Achieving this computational peak, however, dictates a thermodynamic trade-off [Fig.~\ref{Fig_2}(b1, b2)]. The cumulative average irreversible work $\beta W^{\text{irr}}_{\text{tot}}$ upper-bounds the accumulated QID $\chi^{\rm d}_{\rm tot}$, validating the generalized Landauer bound. Both the energetic cost and the informational dissipation exhibit peaks within the critical regions. This reveals that optimal predictive capacity forces the reservoir into a highly susceptible regime that absorbs non-predictive historical data, maximizing informational dissipation and the irreversible heat required for environmental erasure \cite{Mascarenhas_2014_PRE, Varizi_2020_PRR}. 

Finally, the coherence decomposition isolates the quantum origin of this performance enhancement [Fig.~\ref{Fig_2}(c1, c2)]. By choosing the computational basis as the reference dephasing basis, the driving operator $H_1$ is diagonal, ensuring that the fully coherent and classically dephased paradigms incur identical mechanical work. Under this thermodynamic baseline, the ensemble coherences of the memory ($\mathcal{C}^{\rm m}_{\rm tot}$) and predictive ($\mathcal{C}^{\rm p}_{\rm tot}$) states remain positive, confirming the quantum advantage ($\chi^{\rm m}_{\rm tot} \ge \mathcal{I}^{\rm m}_{\rm tot}$ and $\chi^{\rm p}_{\rm tot} \ge \mathcal{I}^{\rm p}_{\rm tot}$) via Eq.~\eqref{eq:holevo_decomposition}. Our simulations verify that $\mathcal{C}^{\rm p}_{\rm tot}$ exceeds $\mathcal{C}^{\rm m}_{\rm tot}$ over broad regimes, demonstrating negative quantum coherent dissipation ($\mathcal{D}^{\rm q}_{\rm tot} < 0$). This mechanism is prominent within the critical regions: while the classical mutual informations ($\mathcal{I}^{\rm m,p}_{\rm tot}$) exhibit localized increases or drops, the ensemble coherences $\mathcal{C}^{\rm m,p}_{\rm tot}$ consistently peak. In addition to classical populations, quantum coherence acts as a physical resource that amplifies the predictive capacity of the many-body reservoir.

\section{Conclusions}
\label{sec:conclusions}

In this work, we establish a non-equilibrium thermodynamic framework to resolve the predictive limits of QRC. Mapping the macroscopic Holevo capacities onto the BKM geometric manifold allows us to analytically evaluate the microscopic mechanisms governing memory retention and target prediction. By introducing QID to quantify the memory redundancy inherent in continuous data processing, we derive a generalized Landauer bound. This bound demonstrates that the energetic penalty of continuous quantum prediction is driven by the thermodynamic cost of erasing retained, non-predictive historical data from the open reservoir.

Applying this framework, we identify the microscopic origin of the optimal computational performance near quantum phase transitions. We show that this peak arises from a spectral resonance rather than merely macroscopic critical fluctuations. As the primary energy gap closes near criticality, the reservoir's internal transition frequencies align with the dominant frequency of the driving signal. The resulting maximization of the dynamic accumulation factor $G(\Delta E_{jk})$, combined with enhanced structural matrix elements, determines the optimal capacity. This mechanism transitions QRC from an empirically optimized architecture to a predictable physical framework. However, this resonance dictates a thermodynamic trade-off: achieving peak utility forces the reservoir to absorb complex history, maximizing informational dissipation and the irreversible entropy production required for environmental erasure. Furthermore, our coherence decomposition demonstrates that off-diagonal quantum coherences amplify predictive capacity without requiring additional mechanical work from the diagonal drive, establishing a resource-efficient quantum advantage.

These results provide theoretical guidance for designing energy-efficient quantum neuromorphic hardware. Achieving optimal processing requires engineering the reservoir's many-body energy spectrum to resonate with the target data's characteristic frequencies, rather than merely tuning to a generic phase transition. This highlights the need for a quantum analogue of the information bottleneck principle \cite{Tishby_2000_arXiv, Murphy_2024_PRL}: future architectures must compress non-predictive history while exploiting coherent resonant phases. Additionally, the critical sensitivity of informational dissipation and predictive capacities offers a paradigm for utilizing machine learning metrics as thermodynamic witnesses for quantum phase transitions \cite{Kobayashi_2025_NC}. Extending this framework to fully quantum data streams, non-ergodic architectures with quantum many-body scars \cite{Bravo_2022_PRXQuantum}, or structured environments with non-Markovian memory will further elucidate the predictive limits of complex quantum information processing \cite{Sannia_2026_npjQI}.

\begin{acknowledgments}

This work was supported by the Shanghai Science and Technology project (24LZ1401600). 

\end{acknowledgments}

\section*{DATA AVAILABILITY}

The data and code that support the findings of this article are openly available \cite{data_availability}.

\appendix
\onecolumngrid

\section{Discrete thermalization map}
\label{sec:appendix_discrete_thermalization_map}

To evaluate the thermodynamic costs of QRC, the open-system dynamics must be treated with theoretical consistency. Local phenomenological master equations, which derive dissipators independently for individual subsystems, fail to capture internal multi-body couplings \cite{Alicki_1976}. 
As shown in Refs.~\onlinecite{Levy_2014, Hofer_2017, Cattaneo_2019, Dann_2021}, such local approaches can violate the second law of thermodynamics by predicting unphysical steady-state heat currents, rendering subsequent energetic analyses invalid. 
To ensure thermodynamic consistency, the open-system dissipator must be microscopically derived from first principles by diagonalizing the instantaneous global system Hamiltonian. 
Applying the standard Born-Markov and secular approximations within this global eigenbasis yields a completely positive master equation that strictly satisfies the Kubo-Martin-Schwinger detailed balance condition \cite{Cattaneo_2019}. 
This guarantees that the open system autonomously relaxes to the exact thermal Gibbs state of the global Hamiltonian, avoiding spurious entropy production \cite{Hofer_2017}. 

To operationalize this thermodynamically consistent relaxation within the discrete-time sequential processing framework, we employ a discrete thermalization map motivated by the repeated-interaction (collision-model) framework \cite{Scarani_2002_PRL, Strasberg_2017_PRX, Ciccarello_2022}. 
The corresponding evolution from time step $t_n$ to $t_{n+1}$ is strictly executed using the CPTP map:
\begin{equation}\label{eq:collisional_map}
	\rho_{t_{n+1}} = \mathcal{E}_{t_n}(\rho_{t_n}) = \left(1 - P_{\text{th}}\right) U_{t_n} \rho_{t_n} U_{t_n}^\dagger + P_{\text{th}} \rho^{\text{eq}}_{s_n}.
\end{equation}
Here, $U_{t_n} = \exp\left(-i H_{t_n} \delta t\right)$ is the exact unitary propagator of the global system Hamiltonian, and the thermalization probability $P_{\text{th}} = 1 - \exp(-\gamma_0 \delta t)$ is determined by the bare thermal relaxation rate $\gamma_0$ and the discrete time step $\delta t$. 
Discrete maps analogous to Eq.~\eqref{eq:collisional_map} are commonly employed in QRC research to capture non-unitary dynamics and memory fading \cite{Nakajima_2017}. 
Crucially, this discrete map perfectly inherits the global thermodynamic consistency established by the continuous-time first-principles derivation \cite{Dann_2021}.  
Because the exact unitary operator strictly commutes with the instantaneous global Gibbs state ($\left[ U_{t_n}, \rho^{\text{eq}}_{s_n} \right] = 0$), the map intrinsically preserves thermal equilibrium, satisfying $\mathcal{E}_{t_n}(\rho^{\text{eq}}_{s_n}) = \rho^{\text{eq}}_{s_n}$. This phenomenological approach rigorously replicates the macroscopic thermodynamic constraints of the microscopic dynamics---guaranteeing complete positivity and correct thermalization---enabling an unambiguous evaluation of the non-equilibrium energetic bounds considered in this work. Furthermore, repeated-interaction schemes of this type are highly suited for implementation on current quantum simulation platforms \cite{Ciccarello_2022}.

\section{Microscopic trajectories and conditional ensembles}
\label{sec:appendix_microscopic_trajectories}

As defined in the main text, the open-system dynamics unravel into an ensemble of stochastic quantum trajectories. The macroscopic conditional operational states---the memory state $\rhomem$ and the predictive state $\rhopred$---are constructed by statistically aggregating the microscopic pure states $|\psi_{t_{n+1}}^{\Gamma_n}\rangle$, conditioned on the present drive $s_n$ and the future target $s_{n+1}$, respectively.

We explicitly demonstrate that both conditional ensembles exactly reconstruct the unconditional average state $\bar{\rho}_{t_{n+1}}$ via probability marginalization. Applying the product rule of probability, $P(s_n) P( S_{n-1}, \Gamma_n | s_n) = P(S_n, \Gamma_n)$, to the ensemble average of the memory states yields:
\begin{equation}\label{eq:memory_marginalization_proof}
\begin{aligned}
	\sum_{s_n} P(s_n) \rhomem &= \sum_{s_n} \sum_{S_{n-1}, \Gamma_n} P(s_n) P\left(S_{n-1}, \Gamma_n | s_n\right) \ket{\psi_{t_{n+1}}^{\Gamma_n}} \!\bra{\psi_{t_{n+1}}^{\Gamma_n}} \\
	&= \sum_{S_n, \Gamma_n} P(S_n, \Gamma_n) \ket{\psi_{t_{n+1}}^{\Gamma_n}}\! \bra{\psi_{t_{n+1}}^{\Gamma_n}} = \bar{\rho}_{t_{n+1}}. 
\end{aligned}
\end{equation}
Similarly, marginalizing over the future target $s_{n+1}$ for the predictive state ensemble yields an analogous consistency identity. Applying the product rule and summing over the target space gives $\sum_{s_{n+1}} P(s_{n+1}) P(S_n, \Gamma_n | s_{n+1}) = \sum_{s_{n+1}} P(S_n, \Gamma_n, s_{n+1}) = P(S_n, \Gamma_n)$. Consequently, the predictive average evaluates to:
\begin{equation}\label{eq:predictive_marginalization_proof}
\begin{aligned}
	\sum_{s_{n+1}} P(s_{n+1}) \rhopred &= \sum_{s_{n+1}} \sum_{S_n, \Gamma_n} P(s_{n+1}) P\left(S_n, \Gamma_n | s_{n+1}\right) \ket{\psi_{t_{n+1}}^{\Gamma_n}}\! \bra{\psi_{t_{n+1}}^{\Gamma_n}} \\
	& = \sum_{S_n, \Gamma_n} P(S_n, \Gamma_n) \ket{\psi_{t_{n+1}}^{\Gamma_n}} \! \bra{\psi_{t_{n+1}}^{\Gamma_n}} = \bar{\rho}_{t_{n+1}}. 
\end{aligned}
\end{equation}
This mathematical consistency guarantees that memory retention and target prediction are evaluated against the exact same physical baseline ensemble. This shared statistical structure is a prerequisite for unambiguously defining the respective Holevo capacities and evaluating the subsequent QID.

\section{Geometric formulation of informational capacities}
\label{sec:geometric_formulation}

To evaluate the thermodynamic limits of QRC, the macroscopic informational capacities---specifically the Holevo quantity---must be formally mapped onto a microscopic geometric manifold. In the weak-driving linear response regime, the BKM metric emerges as the fundamental geometric tensor \cite{Petz_1994, Petz_1996, Rudoy_2009}. 
We evaluate the informational distance between the unconditional average state $\bar{\rho}$ and an infinitesimally perturbed conditional state $\rho = \bar{\rho} + \lambda \delta \rho$, where $\lambda \ll 1$ is a dimensionless scaling parameter and $\delta \rho$ is the state variation operator. Expanding the quantum relative entropy functional $S(\bar{\rho} + \lambda \delta \rho \| \bar{\rho}) = \mathrm{Tr}[(\bar{\rho} + \lambda \delta \rho) \ln (\bar{\rho} + \lambda \delta \rho)] - \mathrm{Tr}[(\bar{\rho} + \lambda \delta \rho) \ln \bar{\rho}]$ order by order in $\lambda$, the zeroth-order term $S(\bar{\rho} \| \bar{\rho})$ naturally vanishes. Because all physical density matrices are trace-normalized [$\mathrm{Tr}(\rho) = \mathrm{Tr}(\bar{\rho}) = 1$], the variation operator must be traceless: $\mathrm{Tr}(\delta \rho) = 0$. Using the cyclic property of the matrix trace, the first-order variation reduces to $\lambda [ \mathrm{Tr}(\delta \rho \ln \bar{\rho}) + \mathrm{Tr}(\delta \rho) - \mathrm{Tr}(\delta \rho \ln \bar{\rho}) ] = \lambda \mathrm{Tr}(\delta \rho)$. Since the perturbation is traceless, this linear term is strictly zero, confirming that the unperturbed state resides at a local minimum. Consequently, the leading non-zero contribution is determined entirely by the second-order expansion, formulated as $\frac{\lambda^2}{2} \Tr [ \delta \rho \left. \partial_\lambda \ln(\bar{\rho} + \lambda \delta \rho) \right|_{\lambda=0}]$.

In classical probability theory, expanding the corresponding classical divergence yields the classical Fisher information metric. In the quantum regime, however, the density matrix $\bar{\rho}$ and its perturbation $\delta \rho$ generally do not commute ($[\bar{\rho}, \delta \rho] \neq 0$). This non-commutativity invalidates standard scalar calculus, requiring the Fréchet derivative of the matrix logarithm to be evaluated via its integral representation:
\begin{equation}
\left. \partial_\lambda \ln(\bar{\rho} + \lambda \delta \rho) \right|_{\lambda=0} = \int_{0}^{\infty} (\bar{\rho} + x I)^{-1} \delta \rho (\bar{\rho} + x I)^{-1} \mathrm{d}x,
\end{equation}
where $I$ is the identity matrix. Substituting this derivative back into the second-order trace expansion yields the BKM metric. As a Riemannian metric tensor on the manifold of quantum states, the BKM metric acts as a bilinear form on the tangent space. Evaluated for a specific tangent vector $\delta \rho$, it is explicitly defined as:
\begin{equation}
g_{\text{BKM}}^{\bar{\rho}}(\delta \rho, \delta \rho) = \int_{0}^{\infty} \mathrm{Tr} \left[ \delta \rho \left(\bar{\rho} + x I\right)^{-1} \delta \rho \left(\bar{\rho} + x I\right)^{-1} \right] \mathrm{d}x.
\label{eq:g_BKM}
\end{equation}
The quantum relative entropy is thus approximated by this quadratic geometric form:
\begin{equation}
S(\bar{\rho} + \lambda \delta \rho \| \bar{\rho}) \approx \frac{\lambda^2}{2} g_{\text{BKM}}^{\bar{\rho}}(\delta \rho, \delta \rho).
\label{eq:S_BKM}
\end{equation}

This geometric formulation establishes the theoretical foundation for our continuous processing framework. The macroscopic memory capacity of the reservoir is quantified by the Holevo quantity $\chi^{\rm m}_{t_{n+1}}$. For an ensemble of conditional memory states $\rho^{\rm m}_{s_n}$ prepared with classical probabilities $P(s_n)$, the unconditional average state is defined as the probability-weighted sum: $\bar{\rho}_{t_{n+1}} = \sum_{s_n} P(s_n) \rho^{\rm m}_{s_n}$ [Eq.~\eqref{eq:memory_marginalization_proof}]. Utilizing this relation alongside the definition of the von Neumann entropy $S(\rho) = -\mathrm{Tr}(\rho \ln \rho)$ and the linearity of the trace, we recast the Holevo capacity as the expected relative entropy between the conditional states and the unconditional average state:
\begin{equation}
	\begin{aligned}
		\chi^{\rm m}_{t_{n+1}} &= S(\bar{\rho}_{t_{n+1}}) - \sum_{s_n} P(s_n) S(\rho^{\rm m}_{s_n}) \\
		&= -\mathrm{Tr} \left[ \left( \sum_{s_n} P(s_n) \rho^{\rm m}_{s_n} \right) \ln \bar{\rho}_{t_{n+1}} \right] + \sum_{s_n} P(s_n) \mathrm{Tr}(\rho^{\rm m}_{s_n} \ln \rho^{\rm m}_{s_n}) \\
		&= \sum_{s_n} P(s_n) \left[ \mathrm{Tr}(\rho^{\rm m}_{s_n} \ln \rho^{\rm m}_{s_n}) - \mathrm{Tr}(\rho^{\rm m}_{s_n} \ln \bar{\rho}_{t_{n+1}}) \right] 
		= \sum_{s_n} P(s_n) S\left( \rho^{\rm m}_{s_n} \| \bar{\rho}_{t_{n+1}} \right).
	\end{aligned}
	\label{eq:expected_relative_entropy}
\end{equation}
This mapping justifies approximating $\chi^{\rm m}_{t_{n+1}}$ via the BKM metric when the conditional states act as weak dynamic perturbations around the average baseline. 
We evaluate these conditional states in the linear response regime. The continuous dynamics are driven by the instantaneous Hamiltonian $H_{t_n} = H_0 + \lambda s_n H_1$, where $H_0$ governs the unperturbed intrinsic many-body dynamics, $H_1$ is a fixed local driving operator of order $\mathcal{O}(1)$, and $\lambda \ll 1$ is the dimensionless weak-coupling parameter. The input stochastic sequence has zero mean ($\sum_{s_n} P(s_n) s_n = 0$) and variance $\sigma_{s_n}^2$. To construct the linear response, we expand the conditional memory state $\rho^{\rm m}_{s_n}(\lambda)$ via a first-order Taylor series around the unperturbed limit $\lambda=0$:
\begin{equation}
	\rho^{\rm m}_{s_n}(\lambda) \approx \rho^{\rm m}_{s_n}(0) + \lambda \left. \frac{\partial \rho^{\rm m}_{s_n}(\lambda)}{\partial \lambda} \right|_{\lambda=0}.
	\label{eq:state_expansion_intermediate}
\end{equation}
Because the weak-coupling parameter $\lambda$ and the classical signal $s_n$ appear strictly as the composite product $x_n \equiv \lambda s_n$ within the driving Hamiltonian, the dynamical state functions fundamentally in terms of this instantaneous coupling, denoted as $\rho(x_n)$. Applying the chain rule to evaluate the derivative in Eq.~\eqref{eq:state_expansion_intermediate} yields:
\begin{equation}
	\frac{\partial \rho^{\rm m}_{s_n}(\lambda)}{\partial \lambda} = \frac{\partial x_n}{\partial \lambda} \frac{\partial \rho(x_n)}{\partial x_n} = s_n \frac{\partial \rho(x_n)}{\partial x_n}.
	\label{eq:chain_rule_derivative}
\end{equation}
This step extracts the classical signal $s_n$ as a linear scaling factor. Evaluated at the unperturbed base point $\lambda=0$ (corresponding to $x_n=0$), the variation operator $\left. \partial_{x_n} \rho(x_n) \right|_{x_n=0}$ is determined entirely by $H_0$ and $H_1$, making it independent of any specific classical realization $s_n$. Consequently, this operator factors out of the statistical summation when averaging Eq.~\eqref{eq:state_expansion_intermediate} over the classical driving:
\begin{equation}
	\bar{\rho}_{t_{n+1}} = \sum_{s_n} P(s_n) \rho^{\rm m}_{s_n}(\lambda) \approx \sum_{s_n} P(s_n) \rho^{\rm m}_{s_n}(0) + \lambda \left( \left. \frac{\partial \rho(x_n)}{\partial x_n} \right|_{x_n=0} \right) \sum_{s_n} P(s_n) s_n.
	\label{eq:state_expansion_unconditional}
\end{equation}
Because the stochastic driving sequence is centered at zero (see Appendix~\ref{subsec:chaotic_generation}), the first-order correction strictly vanishes. This proves that the unperturbed base state coincides with the unconditional average density matrix to first order, yielding $\rho^{\rm m}_{s_n}(0) \approx \bar{\rho}_{t_{n+1}}$. This physical equivalence allows us to formally define the state variation vector $\delta\rho^{\rm m}_{s_n} \equiv \left. \partial_\lambda \rho^{\rm m}_{s_n}(\lambda) \right|_{\lambda=0}$ and rewrite the exact linear response expansion as:
\begin{equation}
	\rho^{\rm m}_{s_n}(\lambda) \approx \bar{\rho}_{t_{n+1}} + \lambda \delta\rho^{\rm m}_{s_n}.
	\label{eq:state_expansion}
\end{equation}
Substituting Eq.~\eqref{eq:state_expansion} directly into the BKM geometric approximation derived in Eq.~\eqref{eq:S_BKM}, the memory capacity translates into a quadratic form. The formal scaling parameter $\lambda$ factors out globally as $\lambda^2$, yielding the final geometric expression:
\begin{equation}
	\chi^{\rm m}_{t_{n+1}} \approx \frac{\lambda^2}{2} \sum_{s_n} P(s_n) g_{\text{BKM}}^{\bar{\rho}_{t_{n+1}}} \left( \delta\rho^{\rm m}_{s_n}, \delta\rho^{\rm m}_{s_n} \right).
	\label{eq:BKM_mem}
\end{equation}

Applying this mathematical framework to evaluate the predictive capacity, the predictive state $\rho^{\rm p}_{s_{n+1}}$ is unraveled from the microscopic trajectories by conditioning strictly on the future target $s_{n+1}$. As proven via probability marginalization [Eq.~\eqref{eq:predictive_marginalization_proof}], this predictive ensemble shares the same unconditional average state $\bar{\rho}_{t_{n+1}}$. Consequently, in the weak-coupling limit, the predictive state admits a parallel first-order Taylor expansion around this shared baseline:
\begin{equation}
	\rho^{\rm p}_{s_{n+1}}(\lambda) \approx \rho^{\rm p}_{s_{n+1}}(0) + \lambda \left. \frac{\partial \rho^{\rm p}_{s_{n+1}}(\lambda)}{\partial \lambda} \right|_{\lambda=0}.
	\label{eq:predictive_state_expansion_intermediate}
\end{equation}
When averaging over the future targets to recover the unconditional state, the first-order correction vanishes due to the zero-mean nature of the target sequence. This confirms that the unperturbed base state coincides with the unconditional average density matrix, yielding $\rho^{\rm p}_{s_{n+1}}(0) \approx \bar{\rho}_{t_{n+1}}$. We formally define the predictive state variation vector as $\delta\rho^{\rm p}_{s_{n+1}} \equiv  \partial_\lambda \rho^{\rm p}_{s_{n+1}}(\lambda) |_{\lambda=0}$. The linear response expansion then simplifies to:
\begin{equation}
	\rho^{\rm p}_{s_{n+1}}(\lambda) \approx \bar{\rho}_{t_{n+1}} + \lambda \delta\rho^{\rm p}_{s_{n+1}}.
	\label{eq:predictive_linear_response}
\end{equation}
Similarly, the predictive capacity $\chi^{\rm p}_{t_{n+1}}$ is equivalent to the expected relative entropy $\sum_{s_{n+1}} P(s_{n+1}) S( \rho^{\rm p}_{s_{n+1}} \| \bar{\rho}_{t_{n+1}})$. Substituting Eq.~\eqref{eq:predictive_linear_response} into the BKM metric expansion [Eq.~\eqref{eq:S_BKM}], we obtain the corresponding quadratic form for the predictive capacity:
\begin{equation}
	\chi^{\rm p}_{t_{n+1}} \approx \frac{\lambda^2}{2} \sum_{s_{n+1}} P(s_{n+1}) g_{\text{BKM}}^{\bar{\rho}_{t_{n+1}}} \left( \delta\rho^{\rm p}_{s_{n+1}}, \delta\rho^{\rm p}_{s_{n+1}} \right).
	\label{eq:final_predictive_identity}
\end{equation}
This mirrors the structure of the memory capacity, demonstrating that both the retention of past inputs and the anticipation of future targets are governed by the same underlying geometric state-space curvature.

\subsection{Analytical capacities for the discrete thermalization map}
\label{subsec:analytical_capacities}

To evaluate the macroscopic informational capacities while preserving the non-Markovian complexity of the classical driving sequence, we perform a direct perturbative expansion on the discrete thermalization map in the weak-coupling limit ($\lambda \to 0$). 
We first establish the statistical consistency of the conditional ensembles. The physical reservoir evolves along a stochastic trajectory parameterized by the classical driving history $S_n = (s_0, s_1, \dots, s_n)$. Let $\rho_{t_{n+1}}(S_n) = \sum_{\Gamma_n} P(\Gamma_n \mid S_n) |\psi_{t_{n+1}}^{\Gamma_n}\rangle\! \langle\psi_{t_{n+1}}^{\Gamma_n}|$ denote the history-dependent microscopic density matrix at time $t_{n+1}$. The macroscopic informational capacities are defined via conditional statistical aggregates. The conditional memory state $\rho^{\rm m}_{s_n}$ and the predictive state $\rho^{\rm p}_{s_{n+1}}$ are constructed by marginalizing over the historical sequences:
\begin{equation}
	\rho^{\rm m}_{s_n} = \sum_{S_{n-1}} P\left( S_{n-1} \mid s_n \right)\rho_{t_{n+1}}(S_n),
	\label{eq:def_memory_state}
\end{equation}
\begin{equation}
	\rho^{\rm p}_{s_{n+1}} = \sum_{S_n} P\left( S_n \mid s_{n+1} \right)\rho_{t_{n+1}}(S_n).
	\label{eq:def_predictive_state}
\end{equation}
To ensure both conditional states share an identical geometric reference [Eqs.~\eqref{eq:memory_marginalization_proof} and \eqref{eq:predictive_marginalization_proof}], we verify their statistical consistency via probability marginalization. For the memory ensemble, multiplying by the marginal probability $P(s_n)$ and summing over the input space yields:
\begin{equation}
	\sum_{s_n} P(s_n) \rho^{\rm m}_{s_n} = \sum_{s_n} P(s_n) \sum_{S_{n-1}} P\left(S_{n-1} \mid s_n\right) \rho_{t_{n+1}}(S_n).
	\label{eq:marginalize_memory_step1}
\end{equation}
Invoking the probability product rule $P(s_n) P\left(S_{n-1} \mid s_n\right) = P(S_{n-1}, s_n) \equiv P(S_n)$, the double summation collapses into the unconditional average state:
\begin{equation}
	\sum_{s_n} P(s_n) \rho^{\rm m}_{s_n} = \sum_{S_n} P(S_n) \rho_{t_{n+1}}(S_n) \equiv \bar{\rho}_{t_{n+1}}.
	\label{eq:marginalize_memory_final}
\end{equation}
Applying an identical procedure to the predictive ensemble produces:
\begin{equation}
	\sum_{s_{n+1}} P(s_{n+1}) \rho^{\rm p}_{s_{n+1}} = \sum_{s_{n+1}} P(s_{n+1}) \sum_{S_n} P\left(S_n \mid s_{n+1}\right) \rho_{t_{n+1}}(S_n).
	\label{eq:marginalize_predictive_step1}
\end{equation}
Because $P(s_{n+1}) P\left(S_n \mid s_{n+1}\right) = P(S_n, s_{n+1})$, summing the joint distribution over all future targets $s_{n+1}$ marginalizes out the future dependence, recovering the historical distribution: $\sum_{s_{n+1}} P(S_n, s_{n+1}) = P(S_n)$. Consequently, the predictive aggregate matches the unconditional baseline:
\begin{equation}
	\sum_{s_{n+1}} P(s_{n+1}) \rho^{\rm p}_{s_{n+1}} = \sum_{S_n} P(S_n) \rho_{t_{n+1}}(S_n) \equiv \bar{\rho}_{t_{n+1}}.
	\label{eq:marginalize_predictive_final}
\end{equation}
This marginalization guarantees that the Holevo relative entropy for both memory retention and target prediction is evaluated against the same unconditional baseline $\bar{\rho}_{t_{n+1}}$ [Eqs.~\eqref{eq:marginalize_memory_final} and \eqref{eq:marginalize_predictive_final}].

As established in Appendix~\ref{sec:appendix_discrete_thermalization_map}, the history-dependent microscopic density matrix evolves iteratively under the discrete thermalization map [Eq.~\eqref{eq:collisional_map}], expressed as $\rho_{t_{n+1}}(S_n) = \mathcal{E}_{t_n}(\rho_{t_n}(S_{n-1}))$. By substituting the instantaneous Hamiltonian $H_{t_n} = H_0 + \lambda s_n H_1$ into the unitary propagator $U_{t_n}$ and the instantaneous Gibbs state $\rho^{\text{eq}}_{s_n}$, the dynamics become explicitly dependent on the weak-coupling parameter $\lambda$. At the zero-driving limit ($\lambda = 0$), the unperturbed steady-state strictly reduces to the intrinsic thermal Gibbs state $\rho_0 = \exp\left(-\beta H_0\right) / Z_0$. 
Because the conditional memory state $\rho^{\rm m}_{s_n}$ and the predictive state $\rho^{\rm p}_{s_{n+1}}$ share the common factor $\rho_{t_{n+1}}(S_n)$ [Eqs.~\eqref{eq:def_memory_state} and \eqref{eq:def_predictive_state}], we first evaluate the linear variation of this history-dependent density matrix with respect to $\lambda$. Applying the product rule yields:
\begin{equation}
	\begin{aligned}
	\left. \partial_\lambda \rho_{t_{n+1}}(S_n) \right|_{\lambda=0} =\, & (1 - P_{\text{th}}) \Big[ (\partial_\lambda U_{t_n}) \rho_{t_n}(S_{n-1}) U_{t_n}^\dagger +  U_{t_n} \rho_{t_n}(S_{n-1}) (\partial_\lambda U_{t_n}^\dagger)  \\
	&  \left.  + U_{t_n} (\partial_\lambda \rho_{t_n}(S_{n-1}))  U_{t_n}^\dagger \Big]\right|_{\lambda=0} 
	+ P_{\text{th}} \left. \partial_\lambda \rho^{\text{eq}}_{s_n} \right|_{\lambda=0}. 
		\end{aligned}
	\label{eq:derivative_product_rule}
\end{equation}
Because $H_1$ generally does not commute with $H_0$, we evaluate the derivatives of the matrix exponentials using the Wilcox integral identity: $\left. \partial_\lambda e^{A + \lambda B} \right|_{\lambda=0} = \int_0^1 e^{A x} B e^{A (1-x)} \dd x$. Applying this to the unitary operator $U_{t_n}$ (with $A = -i H_0 \delta t$ and $B = -i s_n H_1 \delta t$) and performing the substitution $t = x \delta t$ yields $\left. \partial_\lambda U_{t_n} \right|_{\lambda=0} = -i s_n \int_{0}^{\delta t} e^{-i H_0 t} H_1 e^{-i H_0 (\delta t-t)} \dd t$. For the instantaneous Gibbs state $\rho^{\text{eq}}_{s_n}$, applying the quotient rule alongside the Wilcox identity evaluates to $\left. \partial_\lambda \rho^{\text{eq}}_{s_n} \right|_{\lambda=0} = -\beta s_n \int_{0}^{1} \rho_0^y \left(H_1 - \langle H_1 \rangle_0\right) \rho_0^{1-y} \dd y$. Substituting these derivatives and defining the state variation as $\delta\rho_{t_{n+1}}(S_{n}) \equiv \left. \partial_\lambda \rho_{t_{n+1}}(S_n) \right|_{\lambda=0}$, we obtain:
\begin{equation}
	\delta\rho_{t_{n+1}}(S_{n}) =  s_n \mathcal{V} + (1 - P_{\text{th}})\, \mathcal{U} ( \delta\rho_{t_n}(S_{n-1}) ),
	\label{eq:derivative_expansion}
\end{equation}
where the perturbation operator $\mathcal{V}$ and the unperturbed linear map $\mathcal{U}$ are defined as:
\begin{align}
		\mathcal{V} &= -i (1 - P_{\text{th}}) \int_{0}^{\delta t} \left[ e^{-i H_0 t} H_1 e^{i H_0 t}, \rho_0 \right] \dd t  -\beta P_{\text{th}} \int_{0}^{1} \rho_0^y \left(H_1 - \langle H_1 \rangle_0\right) \rho_0^{1-y} \dd y, \\
	 \mathcal{U}(\rho) &= U_0 \rho U_0^\dagger, 
\end{align}
with $\langle H_1 \rangle_0 = \mathrm{Tr}(\rho_0 H_1)$ and $U_0 = \exp\left(-i H_0 \delta t\right)$.

Equation~\eqref{eq:derivative_expansion} defines the recursive relationship between the state variations at consecutive steps. Iteratively unrolling this recursion and utilizing the linearity of the map $\mathcal{U}$ yields:
\begin{equation}
	\delta\rho_{t_{n+1}}(S_n) = s_n \mathcal{V} + (1 - P_{\text{th}})^2 \, \mathcal{U}^2 \left( \delta\rho_{t_{n-1}}(S_{n-2}) \right) + s_{n-1} (1 - P_{\text{th}})\, \mathcal{U}(\mathcal{V}).
	\label{eq:unroll_step_1}
\end{equation}
Repeating this backward substitution down to the initial time step $t_0$ expands the state variation into a sequence of nested unperturbed maps. Because the reservoir is initially prepared in the unperturbed steady state $\rho_0$, which is independent of $\lambda$, the initial variation vanishes ($\delta\rho_{t_0} = 0$). The resulting discrete temporal convolution over all historical perturbations is:
\begin{equation}
	\delta\rho_{t_{n+1}}(S_n) = s_n \mathcal{V} + s_{n-1} (1 - P_{\text{th}}) \, \mathcal{U}(\mathcal{V}) + \dots + s_0 (1 - P_{\text{th}})^n \, \mathcal{U}^n(\mathcal{V}) = \sum_{m=0}^n s_m (1 - P_{\text{th}})^{n-m} \, \mathcal{U}^{n-m}(\mathcal{V}).
	\label{eq:microscopic_variation_unrolled_exact}
\end{equation}
Here, the operator $\mathcal{U}^{n-m}(\mathcal{V})$ represents the physical fading of the historical perturbation $s_m$ after $n-m$ steps of multi-body thermal relaxation. 

From Eq.~\eqref{eq:def_memory_state}, the variation of the conditional memory state evaluates to:
\begin{equation}
\left. \partial_\lambda \rho^{\rm m}_{s_n}(\lambda) \right|_{\lambda=0} = \sum_{S_{n-1}} P\left( S_{n-1} \mid s_n \right) \delta\rho_{t_{n+1}}(S_{n}). 
\label{eq:rhomem_variation}
\end{equation} 
To justify mapping the Holevo capacity onto the BKM metric, the unperturbed baseline state must coincide with the unconditional average state to first order. This requires demonstrating that the statistically averaged state variation vanishes. Utilizing the zero-mean property of the driving sequence, we obtain:
\begin{equation}
	\begin{aligned}
		\sum_{s_{n}} P(s_n) \left. \partial_\lambda \rho^{\rm m}_{s_n}(\lambda) \right|_{\lambda=0} 
		&= \sum_{s_{n}} P(s_n) \sum_{S_{n-1}} P\left( S_{n-1} \mid s_n \right) \delta\rho_{t_{n+1}}(S_{n}) \\
		& = \sum_{S_n} P(S_n) \delta\rho_{t_{n+1}}(S_{n}) 
		= \sum_{S_n} P(S_n) \sum_{m=0}^n s_m (1 - P_{\text{th}})^{n-m} \mathcal{U}^{n-m}(\mathcal{V}) \\
		&= \sum_{m=0}^n \left[ \sum_{S_n} P(S_n) s_m \right] (1 - P_{\text{th}})^{n-m} \mathcal{U}^{n-m}(\mathcal{V}) \\
		&= \sum_{m=0}^n \left[ \sum_{s_m} P(s_m) s_m \right] (1 - P_{\text{th}})^{n-m} \mathcal{U}^{n-m}(\mathcal{V}) = 0.
	\end{aligned}
	\label{eq:variation_average_proof}
\end{equation}
Here, we applied the product rule $P(s_n)P\left(S_{n-1} \mid s_n\right) = P(S_n)$ and marginalized over all historical variables except $s_m$, isolating the statistical mean $\sum_{s_m} P(s_m) s_m = 0$. Substituting Eq.~\eqref{eq:rhomem_variation} into the geometric metric [Eq.~\eqref{eq:BKM_mem}] yields:
\begin{equation}
	\chi^{\rm m}_{t_{n+1}} \approx \frac{\lambda^2}{2} \sum_{s_n} P(s_n) g_{\text{BKM}}^{\bar{\rho}_{t_{n+1}}} \left( \sum_{S_{n-1}} P\left( S_{n-1} \mid s_n \right) \delta\rho_{t_{n+1}}(S_{n}), \sum_{S_{n-1}} P\left( S_{n-1} \mid s_n \right) \delta\rho_{t_{n+1}}(S_{n}) \right).
	\label{eq:BKM_mem_expan}
\end{equation}

We evaluate this metric in the global energy eigenbasis $H_0 \ket{j} = E_j \ket{j}$, with transition energy gaps $\Delta E_{jk} \equiv E_j - E_k$ and unperturbed thermal populations $p_j = \exp\left(-\beta E_j\right)/Z_0$. Evaluating the trace in this basis transforms the metric [Eq.~\eqref{eq:g_BKM}] into a double summation over energy eigenstates:
\begin{equation}
	g_{\text{BKM}}^{\bar{\rho}}(\delta \rho, \delta \rho) = \sum_{j,k} \int_{0}^{\infty} (\delta\rho)_{jk} \frac{1}{p_k + x} (\delta\rho)_{kj} \frac{1}{p_j + x} \dd x.
	\label{eq:bkm_trace_expansion}
\end{equation}
Because the state variation is Hermitian [$(\delta\rho)_{kj} = (\delta\rho)_{jk}^*$], the numerator simplifies to $|(\delta\rho)_{jk}|^2$. The integral over $x$ evaluates analytically via partial fractions:
\begin{equation}
	\int_{0}^{\infty} \frac{1}{\left(p_j + x\right)\left(p_k + x\right)} \dd x = \left[ \frac{1}{p_j - p_k} \ln\left(\frac{p_k + x}{p_j + x}\right) \right]_{0}^{\infty} = \frac{\ln p_j - \ln p_k}{p_j - p_k}.
	\label{eq:bkm_analytic_integral}
\end{equation}
Substituting the thermal populations $p_j = \exp\left(-\beta E_j\right)/Z_0$, the logarithmic difference evaluates to the scaled energy gap: $\ln p_j - \ln p_k = \beta(E_k - E_j)$. Thus, the memory capacity becomes:
\begin{equation}
	\chi^{\rm m}_{t_{n+1}} \approx \frac{\lambda^2}{2} \sum_{s_n} P(s_n) \sum_{j,k} \frac{\beta(E_k - E_j)}{p_j - p_k} \left| \left\langle j \Bigg| \sum_{S_{n-1}} P\left( S_{n-1} \mid s_n \right) \delta\rho_{t_{n+1}}(S_{n}) \Bigg| k\right\rangle \right|^2.
	\label{eq:BKM_mem_expan_ij}
\end{equation}
Substituting Eq.~\eqref{eq:microscopic_variation_unrolled_exact} into this expression yields the explicit analytical form of the memory capacity:
\begin{align}
	\chi^{\rm m}_{t_{n+1}}
	&\approx
	\frac{\lambda^2}{2}  \left[
	\beta
	\sum_{j \neq k} 
	\left( \sum_{s_n} P(s_n)
	\left|
	\sum_{S_{n-1}} P\left( S_{n-1} \mid s_n \right)
	\sum_{m=0}^n s_m g_{jk}^{n-m}
	\right|^2
	\right)
	\frac{p_k - p_j}{E_j - E_k} F_{jk}
	\left| \!\mel{j}{H_1}{k} \right|^2 \right. \nonumber \\
	& \qquad\; \left. + \beta^2 P_{\text{th}}^2 \sum_j
	\left( \sum_{s_n} P(s_n)
	\left|
	\sum_{S_{n-1}} P\left( S_{n-1} \mid s_n \right)
	\sum_{m=0}^n s_m g_{jj}^{n-m}
	\right|^2
	\right)
	 p_j \left| \!\mel{j}{H_1 - \langle H_1 \rangle_0}{j} \right|^2
	\right],
	\label{eq:BKM_mem_final_analytical}
\end{align}
where the effective decay factor is $g_{jk} = (1-P_{\text{th}}) \exp\left(-i \Delta E_{jk} \delta t\right)$ and the dynamical filter is $F_{jk} = |1 - g_{jk}|^2$.

Applying this framework to the predictive capacity, the variation of the conditional predictive state from Eq.~\eqref{eq:def_predictive_state} is $\partial_\lambda \rho^{\rm p}_{s_{n+1}}(\lambda) |_{\lambda=0} = \sum_{S_{n}} P( S_{n} \mid s_{n+1} ) \delta\rho_{t_{n+1}}(S_{n})$. Following an identical geometric expansion in the energy eigenbasis yields the analytical predictive capacity:
\begin{align}
	\chi^{\rm p}_{t_{n+1}}
	&\approx
	\frac{\lambda^2}{2}  \left[
	\beta
	\sum_{j \neq k}
	\left( \sum_{s_{n+1}} P(s_{n+1})
	\left|
	\sum_{S_{n}} P\left( S_{n} \mid s_{n+1} \right)
	\sum_{m=0}^n s_m g_{jk}^{n-m}
	\right|^2
	\right)
	\frac{p_k - p_j}{E_j - E_k} F_{jk}
	\left| \!\mel{j}{H_1}{k} \right|^2 \right. \nonumber \\
	& \qquad\, \left. + \beta^2 P_{\text{th}}^2\sum_j
	\left( \sum_{s_{n+1}} P(s_{n+1})
	\left|
	\sum_{S_{n}} P\left( S_{n} \mid s_{n+1} \right)
	\sum_{m=0}^n s_m g_{jj}^{n-m}
	\right|^2
	\right)
	p_j \left| \!\mel{j}{H_1 - \langle H_1 \rangle_0}{j} \right|^2
	\right].
	\label{eq:BKM_pred_final_analytical}
\end{align}

\begin{figure}
    \centering
    \includegraphics[width=0.93\textwidth]{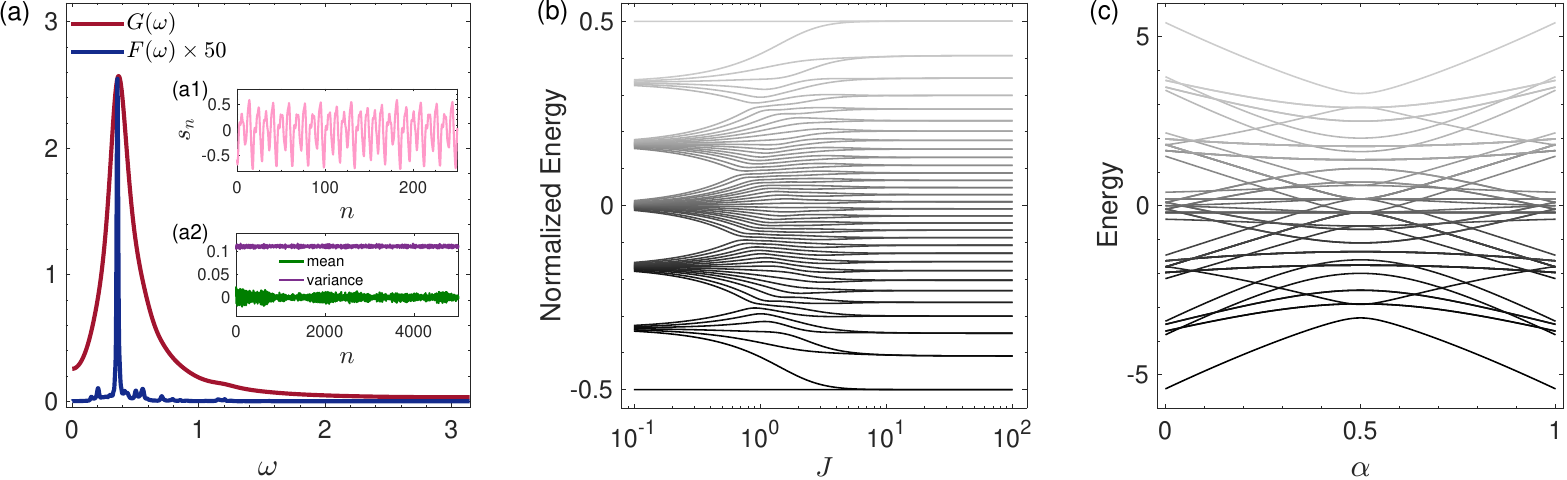}
\caption{
    {\bf Spectral resonance and many-body energy spectra.} 
    (a), Frequency-domain alignment. The temporal accumulation function $G(\omega)$ (red solid line) is calculated using probabilities estimated via the binning method (detailed in Appendix~\ref{sec:numerical_methods}) over 5000 independent time series, each consisting of 5000 discrete time steps. The continuous Fourier power spectrum of the chaotic Mackey-Glass (MG) sequence (blue solid line) is averaged over the same 5000 sequences. Both metrics peak near the characteristic driving frequency $\omega_{\rm s} = 0.36$. Insets: (a1) illustrates a typical segment of the MG sequence $s_n$; (a2) shows the statistical mean (green solid line) and variance (purple solid line) evaluated at each discrete time step $n$ across the 5000 sequences, verifying the zero-mean and constant-variance assumptions. The parameters for the discrete collisional model and the MG time series are identical to those used in the main text. 
    (b), Scaled energy spectrum of the fully connected disordered transverse-field Ising model as a function of the interaction strength $J$. The Hamiltonian and parameters correspond exactly to those defined in the main text. To illustrate the relative state compression and the closing of the energy gap, the energy levels for each $J$ are normalized by the total spectral width. The displayed spectrum is averaged over 10000 random disorder realizations. 
   (c), Unscaled continuous energy spectrum of the augmented cluster model plotted against the tuning parameter $\alpha$. The model configuration and parameters are identical to those utilized in the main text.
    }
    \label{Fig_3}
\end{figure}

\subsection{Microscopic origin of the critical computational peak}
\label{subsec:microscopic_origin}

Empirical observations across various quantum reservoir architectures indicate that the macroscopic information capacities ($\chi^{\rm m}_{t_{n+1}}$ and $\chi^{\rm p}_{t_{n+1}}$) simultaneously maximize within the quantum critical region. To identify the physical mechanisms driving this phenomenon, we analyze the microscopic analytic expressions of the information capacities derived in Eqs.~\eqref{eq:BKM_mem_final_analytical} and \eqref{eq:BKM_pred_final_analytical}. This computational peak arises from an interplay between the dynamic temporal accumulation of the chaotic signal and the divergent structural properties of the reservoir near a continuous quantum phase transition. We detail these physical factors below, focusing explicitly on Eq.~\eqref{eq:BKM_mem_final_analytical}; the analysis for Eq.~\eqref{eq:BKM_pred_final_analytical} is strictly analogous. 

\textit{Dynamic temporal accumulation and spectral resonance.}---The core mechanism governing the temporal memory encoding is the dynamically filtered accumulation factor, defined as:
\begin{equation}
    G(\Delta E_{jk}) = \sum_{s_n} P(s_n)\left| \sum_{S_{n-1}} P( S_{n-1} \mid s_n ) \sum_{m=0}^n s_m g_{jk}^{n-m} \right|^2.
    \label{eq:Gamma_def}
\end{equation}
Here, the factor $g_{jk}^{n-m} = (1-P_{\text{th}})^{n-m} e^{-i \Delta E_{jk} \delta t (n-m)}$ acts as a damped oscillating filter. The inner summation over $m$ functions mathematically as a windowed discrete Fourier transform of the past driving sequence $s_m$, evaluated explicitly at the reservoir's internal energy gap $\Delta E_{jk}$. 

To evaluate $G(\Delta E_{jk})$, we must account for the intrinsic statistical properties of the classical driving sequence. We utilize the MG chaotic time series, which is standardized to reside on a stable chaotic attractor (see Appendix~\ref{subsec:chaotic_generation}). Consequently, the sequence exhibits an approximately zero mean ($\mathbb{E}[s_m] \approx 0$) and a stationary constant variance ($\sigma_{\rm s}^2 = \mathbb{E}[s_m^2]$) [Fig.~\ref{Fig_3}(a2)]. Furthermore, its power spectrum is concentrated at low frequencies, characterized by a dominant peak at a specific characteristic frequency $\omega_{\rm s}$ [empirically $\omega_{\rm s} \approx 0.36$, Fig.~\ref{Fig_3}(a)]. 

By defining the filtered history $Y_{jk}(n) = \sum_{m=0}^n s_m g_{jk}^{n-m}$, the inner summation of Eq.~\eqref{eq:Gamma_def} over the historical trajectories $S_{n-1}$ evaluates the expectation value of $Y_{jk}(n)$ conditioned on the present macroscopic drive $s_n$: $\mathbb{E}_{S_{n-1}}[Y_{jk}(n) \mid s_n] = \sum_{S_{n-1}} P( S_{n-1} \mid s_n ) Y_{jk}(n)$. 
Consequently, the full accumulation factor simplifies to the statistical expectation of the absolute square of this conditional mean: $G(\Delta E_{jk}) = \mathbb{E}_{s_n} [| \mathbb{E}_{S_{n-1}}[Y_{jk}(n) \mid s_n]|^2 ]$. Because the chaotic driving sequence has a zero mean ($\mathbb{E}_{s_m}[s_m] = 0$), the unconditional expectation of the linear sum over the entire history is also strictly zero ($\mathbb{E}_{S_{n}}[Y_{jk}(n)] = 0$). 
According to the law of total expectation, the unconditional expectation of a random variable equals the expected value of its conditional expectation. Thus, averaging $Y_{jk}(n)$ over past histories $S_{n-1}$ for a fixed $s_n$ (yielding $\mathbb{E}_{S_{n-1}}[Y_{jk}(n) \mid s_n]$), and subsequently averaging over all possible states $s_n$, is mathematically identical to averaging $Y_{jk}(n)$ over the entire history $S_n$ simultaneously. Therefore, we obtain $\mathbb{E}_{s_n}\left[ \mathbb{E}_{S_{n-1}}[Y_{jk}(n) \mid s_n] \right] = \mathbb{E}_{S_n}[Y_{jk}(n)] = 0$.

Since the variance of any zero-mean random variable $X$ is defined as its second moment $\mathbb{E}\left[|X|^2\right]$, and we established that the conditional expectation $X = \mathbb{E}_{S_{n-1}}[Y_{jk}(n) \mid s_n]$ has a mean of zero, we can formally identify $G(\Delta E_{jk})$ as the variance of the conditional expectation:
\begin{equation}
    G(\Delta E_{jk}) = \text{Var}_{s_n} \left( \mathbb{E}_{S_{n-1}}[Y_{jk}(n) \mid s_n] \right).
    \label{eq:Gamma_variance_form}
\end{equation}
Evaluating this non-linear conditional expectation exactly for chaotic sequences is generally intractable. However, we approximate it using the linear projection theorem in the Hilbert space of random variables \cite{Moon_2000}. In this space, the inner product between two variables $A$ and $B$ is defined by their covariance $\mathbb{E}_{S_n}[A B^*]$. We approximate the complex non-linear conditional mean $\mathbb{E}_{S_{n-1}}[Y_{jk}(n) \mid s_n]$ with the optimal linear estimator $c \cdot s_n$, where $c$ is a constant complex scalar. 
According to the orthogonal projection theorem, the optimal scalar $c$ minimizing the mean squared error ensures the approximation error is geometrically orthogonal to the basis vector $s_n$. This condition requires the inner product between the error and $s_n$ to be zero: $\mathbb{E}_{S_n} \left[ \left( Y_{jk}(n) - c s_n \right) s_n \right] = 0$. 
Expanding this expectation and isolating $c$ yields $c \cdot \mathbb{E}_{s_n}\left[s_n^2\right] = \mathbb{E}_{S_n}[Y_{jk}(n) s_n]$. Recognizing that $\mathbb{E}_{s_n}\left[s_n^2\right]$ is the constant variance of the input sequence ($\sigma_{\rm s}^2$), the optimal coefficient is $c = \mathbb{E}_{S_n}[Y_{jk}(n) s_n] / \sigma_{\rm s}^2$. This provides the linear approximation for the conditional expectation:
\begin{equation}
    \mathbb{E}_{S_{n-1}}[Y_{jk}(n) \mid s_n] \approx \frac{\mathbb{E}_{S_n}[Y_{jk}(n) s_n]}{\sigma_{\rm s}^2} s_n.
    \label{eq:linear_estimator}
\end{equation} 
Substituting this estimator back into the variance formulation in Eq.~\eqref{eq:Gamma_variance_form} gives:
\begin{equation}
    G(\Delta E_{jk}) \approx \text{Var}_{s_n} \left( \frac{\mathbb{E}_{S_n}[Y_{jk}(n) s_n]}{\sigma_{\rm s}^2} s_n \right).
    \label{eq:Gamma_substituted}
\end{equation}
Because the coefficient $\mathbb{E}_{S_n}[Y_{jk}(n) s_n] / \sigma_{\rm s}^2$ is a deterministic complex scalar, it factors out of the variance operator as its absolute square, leaving the variance of $s_n$ (which is $\sigma_{\rm s}^2$). This simplifies the expression to:
\begin{equation}
    G(\Delta E_{jk}) \approx \left| \frac{\mathbb{E}_{S_n}[Y_{jk}(n) s_n]}{\sigma_{\rm s}^2} \right|^2 \text{Var}(s_n) = \frac{\left| \mathbb{E}_{S_n}[Y_{jk}(n) s_n] \right|^2}{\sigma_{\rm s}^2}.
    \label{eq:Gamma_simplified_fraction}
\end{equation}

The final step is to evaluate the cross-correlation term $\mathbb{E}_{S_n}[Y_{jk}(n) s_n]$. By substituting the explicit definition of $Y_{jk}(n)$ and utilizing the linearity of the expectation operator, we obtain:
\begin{equation}
    \mathbb{E}_{S_n}[Y_{jk}(n) s_n] = \mathbb{E}_{S_n} \left[ \left( \sum_{m=0}^n s_m g_{jk}^{n-m} \right) s_n \right] = \sum_{m=0}^n \mathbb{E}_{s_n, s_m}[s_n s_m] g_{jk}^{n-m}.
    \label{eq:cross_correlation_expansion}
\end{equation}
Introducing the delay variable $\tau = n - m$, and assuming a long, stationary sequence processed beyond the initial transient ($n \to \infty$), the summation extends to infinity, and the expectation $\mathbb{E}_{s_n, s_{n-\tau}}[s_n s_{n-\tau}]$ is precisely the auto-covariance function of the classical input, $C_{\rm a}(\tau)$. Substituting this infinite sum back into Eq.~\eqref{eq:Gamma_simplified_fraction} yields the macroscopic accumulation factor:
\begin{equation}
    G(\Delta E_{jk}) \approx \frac{1}{\sigma_{\rm s}^2} \left| \sum_{\tau=0}^\infty C_{\rm a}(\tau) g_{jk}^\tau \right|^2.
    \label{eq:Gamma_linear_proj}
\end{equation}
To evaluate this factor, we determine $C_{\rm a}(\tau)$. According to the Wiener-Khinchin theorem \cite{Chatfield_1989}, the auto-covariance of a wide-sense stationary random process forms a Fourier transform pair with its power spectral density. Unlike periodic signals with sharp frequency spikes, chaotic systems possess a continuous power spectrum. The macroscopic dynamics are dominated by a characteristic oscillation frequency $\omega_{\rm s}$, but the intrinsic chaotic phase-scrambling broadens this resonant peak into a finite continuous linewidth. 
Modeling this broadened spectral peak using a Lorentzian line shape centered symmetrically at $\pm \omega_{\rm s}$, the inverse discrete-time Fourier transform directly yields an exponentially decaying harmonic oscillation: 
\begin{equation}
    C_{\rm a}(\tau) \approx \sigma_{\rm s}^2 \cos(\omega_{\rm s} \tau\delta t) e^{-\gamma_{\rm s} \tau\delta t}.
    \label{eq:auto_covariance_model}
\end{equation}
Here, the prefactor $\sigma_{\rm s}^2$ ensures $C_{\rm a}(0) = \text{Var}(s_n) = \sigma_{\rm s}^2$. The parameter $\gamma_{\rm s}$ is the half-width at half-maximum of the Lorentzian peak. This finite spectral width $\gamma_{\rm s}$ dictates the intrinsic chaotic decorrelation rate—the inverse timescale over which the continuous signal loses phase coherence and memory. 

Substituting Eq.~\eqref{eq:auto_covariance_model} back into the cross-correlation summation [Eq.~\eqref{eq:cross_correlation_expansion}] yields:
\begin{equation}
    \mathbb{E}_{S_n}[Y_{jk}(n) s_n] \approx \frac{\sigma_{\rm s}^2}{2} \sum_{\tau=0}^\infty \eta^\tau \left[ e^{i(\omega_{\rm s} - \Delta E_{jk})\tau\delta t } + e^{-i(\omega_{\rm s} + \Delta E_{jk})\tau\delta t } \right], 
    \label{eq:cross_corr_expansion}
\end{equation}
where the dimensionless damping factor is $\eta = (1-P_{\text{th}})e^{-\gamma_{\rm s}\delta t}$. 
This expression reveals a strict spectral resonance condition. When the reservoir's internal transition frequency deviates from the classical signal's characteristic frequency ($\Delta E_{jk} \neq \omega_{\rm s}$), the complex phase terms inside the summation oscillate rapidly with the delay $\tau$. Summing over these rapidly rotating phases induces destructive interference, causing the geometric series to collapse toward zero and rendering the information encoding inefficient. 
Conversely, when the reservoir possesses an energy gap that perfectly matches the signal's dominant frequency ($\Delta E_{jk} = \omega_{\rm s}$), the temporal oscillations vanish, leading to constructive interference. 
Under exact resonance, the dominant term evaluates to an infinite geometric series $\sum_{\tau} \eta^\tau = 1 / (1 - \eta)$. Assuming both the thermalization probability and the chaotic decorrelation rate are small per time step, we approximate $1 - \eta \approx P_{\text{th}} + \gamma_{\rm s}\delta t$. Substituting this back into Eq.~\eqref{eq:Gamma_linear_proj}, the accumulation factor reaches its theoretical peak:
\begin{equation}
    G(\omega_{\rm s}) \approx \frac{\sigma_{\rm s}^2}{4(P_{\text{th}} + \gamma_{\rm s}\delta t)^2}.
    \label{eq:Gamma_resonance_peak}
\end{equation}
For our selected parameters ($P_{\text{th}} \approx 0.095$, $\gamma_{\rm s}\delta t \approx 0.015$, and $\sigma_{\rm s}^2 \approx 0.11$), this theoretical maximum evaluates to $G(\omega_{\rm s}) \approx 2.3$, consistent with the empirical peak shown in Fig.~\ref{Fig_3}(a). Crucially, this analytical expression explicitly incorporates the intrinsic chaotic linewidth $\gamma_{\rm s}$. Because chaotic signals inherently decorrelate over time, $\gamma_{\rm s}$ applies a fundamental cutoff to the accumulation process. 

\textit{Thermal regularization and dynamical filtering.}---The thermal regularization factor $\beta(p_k - p_j)/(E_j - E_k)$ naturally favors small energy gaps, reaching its maximum value $\beta p_k$ as $\Delta E_{jk} \to 0$. However, the open-system dynamical filter $F_{jk}$ behaves oppositely. As $\Delta E_{jk} \to 0$, $F_{jk}$ reaches its absolute minimum ($P_{\text{th}}^2$), severely suppressing the capacity near the zero energy gap. This low-frequency suppression is compensated only when the reservoir provides a small energy gap precisely aligning with the classical driving frequency ($\Delta E_{jk} \approx \omega_{\rm s}$) [Eq.~\eqref{eq:Gamma_resonance_peak}].

\textit{Dominance of off-diagonal contributions.}---The spectral resonance condition establishes a distinction between diagonal and off-diagonal contributions to the total capacity. For diagonal matrix elements ($j=k$), the energy gap is exactly zero ($\Delta E_{jj} = 0$). Since the chaotic signal possesses a non-zero characteristic frequency (e.g., $\omega_{\rm s} \approx 0.36$), these elements fail to meet the resonance condition. Their corresponding dynamic accumulation factor, $G(0)$, is suppressed by destructive interference. Consequently, computational capacity is predominantly driven by the off-diagonal elements ($j \neq k$), which provide the finite energy gaps necessary to achieve constructive resonance ($\Delta E_{jk} \approx \omega_{\rm s}$). 

\textit{Structural matrix elements.}---The final components in Eqs.~\eqref{eq:BKM_mem_final_analytical} and \eqref{eq:BKM_pred_final_analytical} are the structural matrix elements determined by the local driving operator $H_1$:
(i) Off-diagonal elements ($| \!\mel{j}{H_1}{k}\!|^2$): These represent the transition probabilities between distinct eigenstates $j$ and $k$ induced by the external field. Near a quantum phase transition, the diverging correlation length and macroscopic delocalization of the many-body eigenstates enhance these matrix elements. This strong coupling provides the primary structural amplification for the capacity.
(ii) Diagonal elements ($| \!\mel{j}{H_1 - \langle H_1 \rangle_0}{j} \!|^2$): This quantity measures the quantum variance of the driving operator within a single eigenstate $j$. Although this intra-state variance can increase near criticality, its macroscopic contribution to the total Holevo capacity remains subdominant because it is paired with the non-resonant and suppressed $G(0)$ factor.

\textit{Synthesis: Physical origin of the peak at the phase transition.}---The computational peak within the quantum critical region arises from the simultaneous occurrence of dynamical spectral resonance and structural amplification. Deep within a fully gapped phase, the intrinsic energy gaps $\Delta E_{jk}$ are significantly larger than the characteristic frequency $\omega_{\rm s}$ of the chaotic drive. The quantum reservoir cannot effectively respond to this low-frequency signal, leading to destructive interference ($G \approx 0$) and near-zero predictive capacity. As the system is tuned toward a continuous quantum phase transition, the primary energy gap closes ($\Delta E \to 0$) and the low-energy spectrum becomes dense [Fig.~\ref{Fig_3}(b,c)]. Because the energy gaps decrease toward zero, the transition frequencies inevitably sweep through the low-frequency domain and match the signal's characteristic frequency $\omega_{\rm s}$. This frequency matching establishes spectral resonance ($\Delta E_{jk} \approx \omega_{\rm s}$). Because this resonance occurs directly at the phase boundary, it coincides with the diverging transition probabilities $| \!\mel{j}{H_1}{k}\! |^2$ of the low-lying off-diagonal states. The combination of these two critical phenomena—constructive temporal accumulation enabled by spectral resonance and structurally enhanced quantum transition probabilities—provides the physical origin for the optimal macroscopic forecasting capacity at the edge of chaos.

\section{Multi-step temporal capacities and scaling within quantum critical regions}
\label{sec:multi_step_capacities}

To evaluate the informational framework across arbitrary temporal scales, we extend our operational definitions to multi-step memory delays ($\tau \ge 0$) and longer-term forecasting horizons ($h \ge 1$). Let $S_n \setminus s_{n-\tau}$ denote the historical driving sequence excluding the specific past input $s_{n-\tau}$. The $\tau$-step delayed memory state $\rho^{\text{m}}(\tau)$ and the $h$-step predictive state $\rho^{\text{p}}(h)$ are constructed from the ensemble of microscopic trajectories:
\begin{equation}\label{eq:extended_rho_mem}
	\rho^{\text{m}}(\tau) = \sum_{S_n \setminus s_{n-\tau}, \Gamma_n} P\left( S_n \setminus s_{n-\tau}, \Gamma_n \mid s_{n-\tau}\right) \ket{\psi_{t_{n+1}}^{\Gamma_n}} \! \bra{\psi_{t_{n+1}}^{\Gamma_n}},
\end{equation}
\begin{equation}\label{eq:extended_rho_pred}
	\rho^{\text{p}}(h) = \sum_{S_n, \Gamma_n} P\left(S_n, \Gamma_n \mid s_{n+h}\right) \ket{\psi_{t_{n+1}}^{\Gamma_n}} \! \bra{\psi_{t_{n+1}}^{\Gamma_n}}.
\end{equation}
To ensure a valid comparison of these capacities, we verify that both conditional ensembles marginalize to the identical unconditional average state $\bar{\rho}_{t_{n+1}}$. For the delayed memory ensemble, applying the product rule of probability yields:
\begin{equation}\label{eq:proof_mem}
	\begin{aligned}
		\sum_{s_{n-\tau}} P(s_{n-\tau}) \rho^{\text{m}}(\tau) &= \sum_{s_{n-\tau}} \, \sum_{S_n \setminus s_{n-\tau}, \Gamma_n} P(s_{n-\tau}) P\left( S_n \setminus s_{n-\tau}, \Gamma_n \mid s_{n-\tau}\right) \ket{\psi_{t_{n+1}}^{\Gamma_n}} \! \bra{\psi_{t_{n+1}}^{\Gamma_n}} \\
		&= \sum_{S_n, \Gamma_n} P(S_n, \Gamma_n) \ket{\psi_{t_{n+1}}^{\Gamma_n}} \! \bra{\psi_{t_{n+1}}^{\Gamma_n}} = \bar{\rho}_{t_{n+1}}.
	\end{aligned}
\end{equation}
Similarly, marginalizing over the future target $s_{n+h}$ in the joint distribution $P(S_n, \Gamma_n, s_{n+h})$ recovers the predictive baseline:
\begin{equation}\label{eq:proof_pred}
	\begin{aligned}
		\sum_{s_{n+h}} P(s_{n+h}) \rho^{\text{p}}(h) &= \sum_{s_{n+h}} \sum_{S_n, \Gamma_n} P(s_{n+h}) P\left(S_n, \Gamma_n \mid s_{n+h}\right) \ket{\psi_{t_{n+1}}^{\Gamma_n}} \! \bra{\psi_{t_{n+1}}^{\Gamma_n}} \\
		&= \sum_{S_n, \Gamma_n} \left[ \sum_{s_{n+h}} P(S_n, \Gamma_n, s_{n+h}) \right] \ket{\psi_{t_{n+1}}^{\Gamma_n}} \! \bra{\psi_{t_{n+1}}^{\Gamma_n}} \\
		&= \sum_{S_n, \Gamma_n} P(S_n, \Gamma_n) \ket{\psi_{t_{n+1}}^{\Gamma_n}} \! \bra{\psi_{t_{n+1}}^{\Gamma_n}} = \bar{\rho}_{t_{n+1}}.
	\end{aligned}
\end{equation}
The instantaneous generalized operational capacities are thus quantified by their respective Holevo quantities:
\begin{equation}\label{eq:extended_chi_mem}
	\chi^{\text{m}}_{t_{n+1}}(\tau) = S(\bar{\rho}_{t_{n+1}}) - \sum_{s_{n-\tau}} P(s_{n-\tau}) S(\rho^{\text{m}}(\tau)),
\end{equation}
\begin{equation}\label{eq:extended_chi_pred}
	\chi^{\text{p}}_{t_{n+1}}(h) = S(\bar{\rho}_{t_{n+1}}) - \sum_{s_{n+h}} P(s_{n+h}) S(\rho^{\text{p}}(h)).
\end{equation}

\begin{figure}
	\centering
	\includegraphics[width=0.93\textwidth]{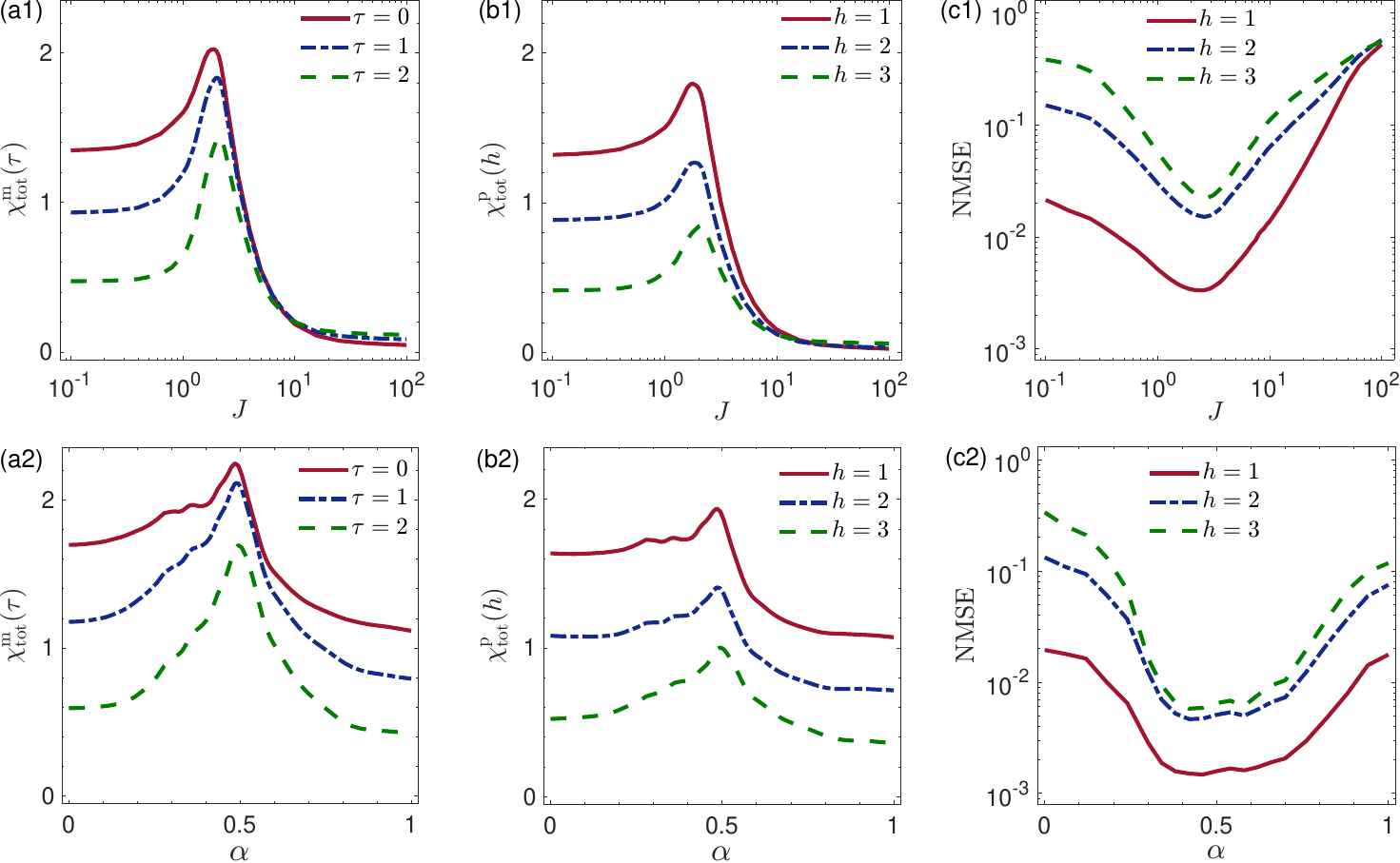}
	\caption{
	{\bf Multi-step temporal capacities and forecasting performance within the quantum critical regions.} 
	The top row (a1-c1) presents results for the disordered TFIM, while the bottom row (a2-c2) corresponds to the augmented cluster model. 
	(a) The generalized quantum memory capacity $\chi^{\text{m}}_{\rm tot}(\tau)$ as a function of the respective tuning parameter ($J$ or $\alpha$) for delays $\tau = 0, 1, 2$. 
	(b) The quantum predictive capacity $\chi^{\text{p}}_{\rm tot}(h)$ for forecasting horizons $h = 1, 2, 3$. 
	(c) The empirical normalized mean squared error (NMSE) on a logarithmic scale for identical forecasting horizons $h = 1, 2, 3$. All reservoir configurations and physical simulation parameters are strictly identical to those described in Fig.~\ref{Fig_2} of the main text.
	}
	\label{Fig_4}
\end{figure}

To evaluate the macroscopic capacities over a finite sequence, we sum the instantaneous metrics. The summation boundaries are truncated to preserve causality and prevent out-of-domain indexing:
\begin{equation}\label{eq:summed_capacities}
	\chi^{\text{m}}_{\rm tot}(\tau) = \sum_{n=\tau}^{N-1} \chi^{\text{m}}_{t_{n+1}}(\tau), \quad \text{and} \quad \chi^{\text{p}}_{\rm tot}(h) = \sum_{n=0}^{N-h} \chi^{\text{p}}_{t_{n+1}}(h).
\end{equation}
The lower bound $n=\tau$ ensures the existence of historical drives $s_{n-\tau} \ge s_0$, and the upper bound $N-h$ guarantees the forecasting targets $s_{n+h} \le s_{N}$ remain within the available time-series. 
We validate these extended metrics using the chaotic MG sequence, applying the same many-body architectures and physical parameters defined in Fig.~\ref{Fig_2} of the main text. As shown in Fig.~\ref{Fig_4}(a) and (b), both the generalized memory capacity $\chi^{\text{m}}_{\rm tot}(\tau)$ and the predictive capacity $\chi^{\text{p}}_{\rm tot}(h)$ exhibit pronounced peaks within their respective quantum critical regions. Due to the fading memory property of dissipative open dynamics, higher-order delays ($\tau=1, 2$) and longer prediction horizons ($h=2, 3$) systematically reduce the absolute extractable information. Across all temporal scales, these informational maxima perfectly align with the global minima of the forecasting error [NMSE, Fig.~\ref{Fig_4}(c)]. This alignment confirms that the optimal macroscopic task performance within the quantum critical region is directly governed by the maximization of the underlying microscopic multi-step information capacities.

\section{Coherence decomposition and Markovian bounds on coherent dissipation}
\label{sec:appendix_coherence}

We detail the bipartition of the Holevo quantity $\chi = \mathcal{I} + \mathcal{C}$ \cite{Lecamwasam_2024_PRXQuantum}. For a generic quantum ensemble encoding a classical variable $X$ into conditional states $\rho_x$ with probabilities $P(x)$, the extractable Holevo information is defined as \cite{Holevo_1973, Schumacher_1997_PRA, Holevo_1998, Holevo_2001_Book, Vedral_2002_RMP, Nielsen_Chuang_2010, Wilde_2017}: $\chi = S(\bar{\rho}) - \sum_x P(x) S(\rho_x)$, 
where $\bar{\rho} = \sum_x P(x) \rho_x$ is the unconditional average density matrix. Rearranging the relative entropy of coherence $C(\rho) = S[\Delta(\rho)] - S(\rho)$ \cite{Baumgratz_2014_PRL, Streltsov_2017_RMP}, the von Neumann entropy of an arbitrary state expands as $S(\rho) = S[\Delta(\rho)] - C(\rho)$. We thus have:
\begin{equation}\label{eq:holevo_substitution_step}
	\chi = \left( S[\Delta(\bar{\rho})] - C(\bar{\rho}) \right) - \sum_x P(x) \left( S[\Delta(\rho_x)] - C(\rho_x) \right).
\end{equation}
Grouping the population and coherence terms results in:
\begin{equation}\label{eq:holevo_grouped_step}
	\chi = \left( S[\Delta(\bar{\rho})] - \sum_x P(x) S[\Delta(\rho_x)] \right) + \left( \sum_x P(x) C(\rho_x) - C(\bar{\rho}) \right).
\end{equation}
Because the completely dephasing channel $\Delta$ is a linear trace-preserving map, the dephased average state satisfies $\Delta(\bar{\rho}) = \sum_x P(x) \Delta(\rho_x)$. Consequently, the first bracketed term defines the classical Shannon mutual information, $\mathcal{I} = S[\Delta(\bar{\rho})] - \sum_x P(x) S[\Delta(\rho_x)]$, which quantifies the information extractable exclusively via the diagonal populations \cite{Wilde_2017}. Recognizing the second term as the ensemble coherence $\mathcal{C}$, Eq.~\eqref{eq:holevo_grouped_step} establishes the information bipartition $\chi = \mathcal{I} + \mathcal{C}$.

Applying this formulation to our continuous temporal processing framework partitions the total QID into classical population dissipation and quantum coherent dissipation, $\chidiss = \mathcal{D}^{\rm c}_{t_{n+1}} + \mathcal{D}^{\rm q}_{t_{n+1}}$. The classical population dissipation evaluates to $\mathcal{D}^{\rm c}_{t_{n+1}} = \mathcal{I}^{\rm m}_{t_{n+1}} - \mathcal{I}^{\rm p}_{t_{n+1}}$. Because the memory and predictive capacities are evaluated over the same unconditional average state $\bar{\rho}_{t_{n+1}}$ [as demonstrated in Eqs.~\eqref{eq:memory_marginalization_proof} and \eqref{eq:predictive_marginalization_proof}], the global coherence term $C(\bar{\rho}_{t_{n+1}})$ cancels out during the subtraction. This isolates the coherent dissipation, which quantifies the dynamic coherences mobilized to encode the past that fail to correlate with the predictive target:
\begin{equation}\label{eq:coherent_dissipation_definition}
	\mathcal{D}^{\rm q}_{t_{n+1}} = \sum_{s_n} P(s_n) C (\rhomem ) - \sum_{s_{n+1}} P(s_{n+1}) C(\rhopred ).
\end{equation}
This coherent dissipation is non-negative ($\mathcal{D}^{\rm q}_{t_{n+1}} \ge 0$) when the classical driving sequence is a Markov process. In fact, under Markovian driving, the transition to the future target $s_{n+1}$ depends exclusively on the present drive $s_n$. Because this external classical drive is conditionally independent of the historical sequence $S_{n-1}$ and the reservoir's internal quantum jumps $\Gamma_n$, the joint probability factorizes as $P(S_n, \Gamma_n \mid s_{n+1}) = P(s_n \mid s_{n+1}) P(S_{n-1}, \Gamma_n \mid s_n)$. Substituting this factorization into the definition of the predictive state allows $\rhopred$ to be reconstructed as a convex causal mixture of the memory states $\rhomem$:
\begin{equation}\label{eq:predictive_causal_mixture}
	\rhopred = \sum_{s_n} P(s_n \mid s_{n+1}) \rhomem.
\end{equation}
Because the quantum relative entropy $S(\rho \,\|\, \sigma)$ is jointly convex \cite{Wilde_2017} and the dephasing map $\Delta$ is a linear CPTP channel, the coherence measure $C(\rho)$ is convex \cite{Baumgratz_2014_PRL}. Applying this convexity property to the causal mixture structure in Eq.~\eqref{eq:predictive_causal_mixture} yields:
\begin{equation}\label{eq:convexity_of_predictive_coherence}
	C(\rhopred) \le \sum_{s_n} P(s_n \mid s_{n+1}) C(\rhomem).
\end{equation}
Multiplying both sides of Eq.~\eqref{eq:convexity_of_predictive_coherence} by the marginal probability $P(s_{n+1})$ and summing over all future targets yields:
\begin{equation}\label{eq:summed_coherence_inequality_step1}
	\sum_{s_{n+1}} P(s_{n+1}) C(\rhopred) \le \sum_{s_{n+1}} P(s_{n+1}) \sum_{s_n} P(s_n \mid s_{n+1}) C(\rhomem).
\end{equation}
Using the product rule of probability $P(s_{n+1}) P(s_n \mid s_{n+1}) = P(s_n, s_{n+1})$, the right-hand side of Eq.~\eqref{eq:summed_coherence_inequality_step1} consolidates into a joint expectation:
\begin{equation}\label{eq:consolidated_coherence_joint_inequality}
	\sum_{s_{n+1}} P(s_{n+1}) C(\rhopred) \le \sum_{s_n, s_{n+1}} P(s_n, s_{n+1}) C(\rhomem).
\end{equation}
Summing the joint probability distribution over all mutually exclusive future target states eliminates the future temporal dependence, recovering the marginal probability of the present state, $\sum_{s_{n+1}} P(s_n, s_{n+1}) = P(s_n)$. Substituting this marginalization into the right-hand side of Eq.~\eqref{eq:consolidated_coherence_joint_inequality} establishes the bounding coherence inequality:
\begin{equation}\label{eq:summed_coherence_inequality}
	\sum_{s_{n+1}} P(s_{n+1}) C(\rhopred) \le \sum_{s_n} P(s_n) C(\rhomem).
\end{equation}
Subtracting the left-hand side of Eq.~\eqref{eq:summed_coherence_inequality} from the right-hand side proves that $\mathcal{D}^{\rm q}_{t_{n+1}} \ge 0$ for Markovian driving. 

Conversely, for non-Markovian driving sequences, this probability factorization breaks down, invalidating the convex causal mixture structure of Eq.~\eqref{eq:predictive_causal_mixture}. This allows $\mathcal{D}^{\rm q}_{t_{n+1}} < 0$. As established in the main text, rather than directly suppressing physical heat, this negative coherent dissipation lowers the theoretical minimum of the generalized Landauer bound. This formalizes the thermodynamic quantum advantage, demonstrating that off-diagonal quantum interference serves as a physical resource to relax the theoretical threshold for irreversible environmental dissipation.

\section{Derivation of the generalized Landauer bound}
\label{sec:appendix_landauer}

We detail the derivation of the generalized Landauer bound. 
As shown in the main text, the macroscopic non-equilibrium free energy functional is defined as $\mathcal{F}(\rho) = \Tr(\rho H) - \beta^{-1} S(\rho)$ \cite{Gaveau_1997, Sivak_2012_PRL, Parrondo_2015_NP}. During the temporal information injection step at $t_{n+1}$, the input scalar is updated via a sudden quench. This shifts the Hamiltonian from $H_{t_n} = H_0 + s_{n}\lambda H_1$ to $H_{t_{n+1}} = H_0 + s_{n+1}\lambda H_1$, while the microscopic pure state $|\psi_{t_{n+1}}^{\Gamma_n}\rangle$ remains dynamically frozen. 
The microscopic stochastic work $w_{n+1}$ performed on the reservoir equals the energy differential of the driving field: $w_{n+1} = \langle\psi_{t_{n+1}}^{\Gamma_n}|H_{t_{n+1}}|\psi_{t_{n+1}}^{\Gamma_n}\rangle - \langle\psi_{t_{n+1}}^{\Gamma_n}|H_{t_n}|\psi_{t_{n+1}}^{\Gamma_n}\rangle$ \cite{Elouard_2017_npjQI, Alonso_2016_PRL, Strasberg_2019_PRE}. The macroscopic average work $W_{n+1}$ is the statistical expectation of this microscopic work over all classical input sequences and quantum jump trajectories. Expanding the post-quench and pre-quench energy expectation values for $W_{n+1} = \sum_{S_{n+1}, \Gamma_{n}} P(S_{n+1}, \Gamma_{n}) w_{n+1}$ yields:
\begin{align}
	W_{n+1} &= \sum_{S_{n+1}, \Gamma_n} P(S_{n+1}, \Gamma_n) \left[ \langle\psi_{t_{n+1}}^{\Gamma_n}|H_{t_{n+1}}|\psi_{t_{n+1}}^{\Gamma_n}\rangle - \langle\psi_{t_{n+1}}^{\Gamma_n}|H_{t_n}|\psi_{t_{n+1}}^{\Gamma_n}\rangle \right] \nonumber\\
	&= \sum_{s_{n+1}} P(s_{n+1}) \Tr \left[ \left( \sum_{S_n, \Gamma_n} P(S_n, \Gamma_n \mid s_{n+1}) \ket{\psi_{t_{n+1}}^{\Gamma_n}} \! \bra{\psi_{t_{n+1}}^{\Gamma_n}} \right) H_{t_{n+1}} \right] \nonumber \\
	&\quad - \sum_{s_n} P(s_n) \Tr \left[ \left( \sum_{S_{n-1}, \Gamma_n} P(S_{n-1}, \Gamma_n \mid s_n) \ket{\psi_{t_{n+1}}^{\Gamma_n}} \! \bra{\psi_{t_{n+1}}^{\Gamma_n}} \right) H_{t_n} \right] \nonumber \\
	&= \sum_{s_{n+1}} P(s_{n+1}) \Tr \left( \rhopred H_{t_{n+1}} \right) - \sum_{s_n} P(s_n) \Tr \left( \rhomem H_{t_n} \right). \label{eq:average_work_expanded}
\end{align}
To derive this macroscopic form, we apply the product rule of probability to condition the joint distributions on the instantaneous inputs. Marginalizing over the conditionally independent historical trajectories reconstructs the traces over the predictive state $\rhopred$ and the memory state $\rhomem$. 

The instantaneous average irreversible work during the information injection stage is the deviation of the actual average work from the macroscopic free energy difference \cite{Parrondo_2015_NP}: $W^{\text{irr}}_{n+1} = W_{n+1} - \Delta \mathcal{F}_{n+1}$. Substituting the non-equilibrium free energy definition into this difference yields:
\begin{align}
	\beta \Delta \mathcal{F}_{n+1} &= \sum_{s_{n+1}} P(s_{n+1}) \beta \left[ \Tr(\rhopred H_{t_{n+1}}) - \beta^{-1} S(\rhopred) \right] - \sum_{s_n} P(s_n) \beta \left[ \Tr(\rhomem H_{t_n}) - \beta^{-1} S(\rhomem) \right] \nonumber \\
	&= \beta \left[ \sum_{s_{n+1}} P(s_{n+1}) \Tr( \rhopred H_{t_{n+1}} ) - \sum_{s_n} P(s_n) \Tr( \rhomem H_{t_n} ) \right] \nonumber \\
	&\quad - \left[ \sum_{s_{n+1}} P(s_{n+1}) S(\rhopred) - \sum_{s_n} P(s_n) S(\rhomem) \right] \nonumber \\
	&= \beta W_{n+1} - \Delta S_{n+1}.
\end{align}
The energetic expectations cancel the mechanical work, demonstrating that the instantaneous dissipation arises entirely from the entropic penalty: $\beta W^{\text{irr}}_{n+1} = \beta W_{n+1} - \beta \Delta \mathcal{F}_{n+1} = \Delta S_{n+1}$. Applying the Holevo quantity definition to these conditional ensembles translates this entropic cost directly into the QID:
\begin{equation}\label{eq:work_dissipation_equivalence_si}
\begin{aligned}
	\beta W^{\text{irr}}_{n+1} &= \sum_{s_{n+1}} P(s_{n+1}) S(\rhopred) - \sum_{s_n} P(s_n) S(\rhomem) \\
	&= \left[ S(\bar{\rho}_{t_{n+1}}) - \chipred \right] - \left[ S(\bar{\rho}_{t_{n+1}}) - \chimem \right] = \chidiss.
\end{aligned}
\end{equation}

Following the sudden quench, the reservoir undergoes a thermalization interval of duration $\delta t$ governed by the fixed Hamiltonian $H_{t_{n+1}}$. Because the Hamiltonian remains constant, the microscopic quantum heat $q_{n+1}$ exchanged with the bath equals the change in the system's energy during this interval: $q_{n+1} = \langle\psi_{t_{n+2}}^{\Gamma_{n+1}}|H_{t_{n+1}}|\psi_{t_{n+2}}^{\Gamma_{n+1}}\rangle - \langle\psi_{t_{n+1}}^{\Gamma_n}|H_{t_{n+1}}|\psi_{t_{n+1}}^{\Gamma_n}\rangle$. Averaging over the statistical ensemble yields the macroscopic heat $Q_{n+1} = \sum_{S_{n+1}, \Gamma_{n+1}} P(S_{n+1}, \Gamma_{n+1}) q_{n+1}$. During this relaxation phase, the relevant conditional ensemble evolves from the predictive state $\rhopred$ to the subsequent memory state $\rho^{\rm m}_{s_{n+1}}$. Applying the CPTP map $\mathcal{E}_{t_{n+1}}$ to the predictive ensemble verifies this evolution:
\begin{align}
	\mathcal{E}_{t_{n+1}}(\rhopred) &= \sum_{S_n, \Gamma_n} P(S_n, \Gamma_n \mid s_{n+1}) \sum_{\gamma_{n+1}} M_{\gamma_{n+1}}(t_{n+1}) \ket{\psi_{t_{n+1}}^{\Gamma_n}} \! \bra{\psi_{t_{n+1}}^{\Gamma_n}} M_{\gamma_{n+1}}^\dagger(t_{n+1}) \nonumber \\
	&= \sum_{S_n, \Gamma_n} P(S_n, \Gamma_n \mid s_{n+1}) \sum_{\gamma_{n+1}} P(\gamma_{n+1} \mid S_{n+1}, \Gamma_n) \ket{\psi_{t_{n+2}}^{\Gamma_{n+1}}}\bra{\psi_{t_{n+2}}^{\Gamma_{n+1}}}.
\end{align}
Applying the probability chain rule, $P(S_n, \Gamma_n \mid s_{n+1}) P(\gamma_{n+1} \mid S_{n+1}, \Gamma_n) = P(S_n, \Gamma_{n+1} \mid s_{n+1})$, we contract the summation over the complete quantum jump trajectory:
\begin{align}
	\mathcal{E}_{t_{n+1}}(\rhopred) &= \sum_{S_n, \Gamma_{n+1}} P(S_n, \Gamma_{n+1} \mid s_{n+1}) \ket{\psi_{t_{n+2}}^{\Gamma_{n+1}}}\bra{\psi_{t_{n+2}}^{\Gamma_{n+1}}} \equiv \rho^{\rm m}_{s_{n+1}}.
\end{align}
This demonstrates that the continuous dynamical map reconstructs the memory state at the next temporal node.

Because the Hamiltonian remains constant during this relaxation phase, no external mechanical work is performed. The average irreversible work produced during this step, $W^{\text{relax}}_{n+1} = - \Delta \mathcal{F}^{\text{relax}}_{n+1}$, originates from the non-unitary spontaneous relaxation toward the instantaneous global thermal Gibbs state $\rho^{\text{eq}}_{s_{n+1}} = \exp(-\beta H_{t_{n+1}})/ \Tr[\exp(-\beta H_{t_{n+1}})]$. The macroscopic change in free energy is the statistical average over all future inputs $s_{n+1}$:
\begin{align}
	\Delta \mathcal{F}^{\text{relax}}_{n+1} &= \sum_{s_{n+1}} P(s_{n+1}) \left( \mathcal{F}[\rho^{\rm m}_{s_{n+1}}] - \mathcal{F}[\rhopred] \right) \nonumber \\
	&= \beta^{-1} \sum_{s_{n+1}} P(s_{n+1}) \left[ S(\rho^{\rm m}_{s_{n+1}} \,\|\, \rho^{\text{eq}}_{s_{n+1}}) - S(\rhopred \,\|\, \rho^{\text{eq}}_{s_{n+1}}) \right].
\end{align}
Due to the monotonicity of the quantum relative entropy under CPTP maps [$S(\mathcal{E}(\rho) \,\|\, \mathcal{E}(\sigma)) \le S(\rho \,\|\, \sigma)$] \cite{Wilde_2017}, the informational divergence from the invariant Gibbs state cannot increase. Therefore, the dissipation during thermal relaxation is non-negative:
\begin{equation}\label{eq:relaxation_irreversible_work}
	\beta W^{\text{relax}}_{n+1} = \sum_{s_{n+1}} P(s_{n+1}) \left[ S(\rhopred \,\|\, \rho^{\text{eq}}_{s_{n+1}}) - S(\rho_{s_{n+1}}^{\rm m} \,\|\, \rho^{\text{eq}}_{s_{n+1}}) \right] \ge 0.
\end{equation}
Summing the energetic contributions of both the sudden signal injection and the thermal relaxation phases over the entire computational sequence of duration $N$ yields the total accumulated irreversible work $W^{\text{irr}}_{\text{tot}}$. Because the relaxation contributions are non-negative ($W^{\text{relax}}_{n+1} \ge 0$), discarding them establishes a macroscopic inequality bounded by the informational dissipation:
\begin{equation}\label{eq:total_work_bound_si}
	\beta W^{\text{irr}}_{\text{tot}} = \sum_{n=0}^{N-1} \beta \left( W^{\text{irr}}_{n+1} + W^{\text{relax}}_{n+1} \right) \ge \sum_{n=0}^{N-1} \chidiss.
\end{equation}

To derive the generalized Landauer bound governing this continuous process, we apply the macroscopic first law of thermodynamics: $\Delta U_{\text{tot}} = W_{\text{tot}} + \sum_{n=0}^{N-1} Q_{n+1} = W_{\text{tot}} - Q_{\text{diss}}$. Here, $W_{\text{tot}} = \sum_{n=0}^{N-1} W_{n+1}$ is the total average work performed on the system, and $Q_{\text{diss}}$ is the total heat dissipated into the thermal bath. The total irreversible work is defined relative to the macroscopic non-equilibrium free energy difference:
\begin{align}
	\beta W^{\text{irr}}_{\text{tot}} &= \beta W_{\text{tot}} - \beta \Delta \mathcal{F}_{\text{tot}} = \beta W_{\text{tot}} - \beta \left( \Delta U_{\text{tot}} + \beta^{-1} \Delta S_{\text{sys}} \right) = \beta Q_{\text{diss}} - \Delta S_{\text{sys}}.
	\label{eq:first_law_irreversible_identity}
\end{align}
Here, $\Delta S_{\text{sys}} = \sum_{s_0} P(s_0) S(\rho^{\rm m}_{s_0}) - \sum_{s_N} P(s_N) S(\rho^{\rm m}_{s_{N}})$ is the net change in the conditional von Neumann entropy of the reservoir between the initial and final states. Substituting the identity from Eq.~\eqref{eq:first_law_irreversible_identity} into Eq.~\eqref{eq:total_work_bound_si}, and replacing $\chidiss$ with its coherence decomposition $\chidiss = \mathcal{D}^{\rm c}_{t_{n+1}} + \mathcal{D}^{\rm q}_{t_{n+1}}$, yields the coherent Landauer limit:
\begin{equation}\label{eq:generalized_coherent_landauer_bound}
	\beta Q_{\text{diss}} \ge \Delta S_{\text{sys}} + \sum_{n=0}^{N-1} \chidiss = \Delta S_{\text{sys}} + \sum_{n=0}^{N-1} \left( \mathcal{D}^{\rm c}_{t_{n+1}} + \mathcal{D}^{\rm q}_{t_{n+1}} \right).
\end{equation}
To isolate the thermodynamic consequences of quantum interference, we evaluate the boundary entropy change $\Delta S_{\text{sys}}$. Applying the dephasing identity $S(\rho) = S[\Delta(\rho)] - C(\rho)$ to each conditional state, we decompose this boundary term into classical population and quantum coherence contributions:
\begin{equation}
	\Delta S_{\text{sys}} = \Delta S_{\text{sys}}^{\rm c} - \Delta C_{\text{sys}}.
\end{equation}
Here, $\Delta S_{\text{sys}}^{\rm c} = \sum_{s_0} P(s_0) S[\Delta(\rho^{\rm m}_{s_0})] - \sum_{s_N} P(s_N) S[\Delta(\rho^{\rm m}_{s_{N}})]$ represents the net change in the conditional Shannon entropy of the diagonal populations (the classical state-compression cost), and $\Delta C_{\text{sys}} = \sum_{s_0} P(s_0) C(\rho^{\rm m}_{s_0}) - \sum_{s_N} P(s_N) C(\rho^{\rm m}_{s_{N}})$ represents the corresponding change in the ensemble coherence. 
Substituting this boundary decomposition back into Eq.~\eqref{eq:generalized_coherent_landauer_bound}, we reorganize the generalized Landauer bound into two physically distinct components \cite{Van_2022_PRL}:
\begin{equation}\label{eq:ultimate_coherent_bound}
	\beta Q_{\text{diss}} \ge \underbrace{ \left[\Delta S_{\text{sys}}^{\rm c} + \sum_{n=0}^{N-1} \mathcal{D}^{\rm c}_{t_{n+1}} \right] }_{\mathcal{L}^{\rm c}} + \underbrace{ \left[ - \Delta C_{\text{sys}} + \sum_{n=0}^{N-1} \mathcal{D}^{\rm q}_{t_{n+1}} \right] }_{\mathcal{L}^{\rm q}}.
\end{equation}
This separation demonstrates that Landauer's original quasi-static erasure bound ($\Delta S_{\text{sys}}^{\rm c}$) is insufficient to describe continuous temporal sequence processing. The total heat dissipated into the environment is governed by a classical population bound ($\mathcal{L}^{\rm c}$)---driven by the continuous non-equilibrium penalty of retaining non-predictive historical populations---and a quantum coherent contribution ($\mathcal{L}^{\rm q}$). 
Crucially, the coherent contribution $\mathcal{L}^{\rm q}$ dictates whether quantum interference manifests as an additional thermodynamic penalty or an energetic advantage. If the reservoir mobilizes quantum superpositions to encode historical features that provide no statistical utility for forecasting ($\mathcal{L}^{\rm q} > 0$), environmental decoherence collapses these misaligned off-diagonal structures, increasing the threshold for heat dissipation. Conversely, when processing complex non-Markovian sequences where dynamic coherences successfully align with the predictive target ($\mathcal{L}^{\rm q} < 0$), off-diagonal quantum interference lowers the thermodynamic bound. In this advantageous regime, the total heat dissipation threshold is relaxed below the classical population limit $\mathcal{L}^{\rm c}$, opening a thermodynamic window for enhanced energetic efficiency.

\section{Numerical methods and simulation details}
\label{sec:numerical_methods}

\subsection{Chaotic time-series generation}
\label{subsec:chaotic_generation}

The continuous classical input sequence is generated using the MG delay differential equation \cite{MG_1977}:
\begin{equation}\label{eq:mackey_glass}
	\frac{\dd s(t)}{\dd t} = \frac{\beta_{\mathrm{MG}}\,s(t-\tau_{\mathrm{MG}})}{1+s^{10}(t-\tau_{\mathrm{MG}})} - \gamma_{\mathrm{MG}}\,s(t). 
\end{equation}
Originally introduced to model nonlinear physiological feedback systems, the MG equation has become a standard benchmark for time-series prediction tasks. Throughout this work, we set $\beta_{\mathrm{MG}}=0.2$, $\gamma_{\mathrm{MG}}=0.1$, and $\tau_{\mathrm{MG}}=18$ to ensure chaotic dynamics \cite{Nakajima_2017}. 

The equation is numerically integrated using a fourth-order Runge-Kutta method. After discarding initial transients to ensure the trajectory converges to the chaotic attractor, the solution is sampled at a temporal interval of $\delta t_{\mathrm{samp}}=3$ to obtain the discrete sequence $s(n\delta t_{\mathrm{samp}})$. These sampled data are then linearly rescaled to the interval $[-1,1]$ to produce the driving sequence $\{s_n\}$ injected into the quantum reservoir. The resulting sequence is approximately stationary over the sampled time window, characterized by a near-zero mean and a time-invariant variance. Its Fourier power spectrum exhibits a broadband continuous distribution with a pronounced low-frequency peak, reflecting the coexistence of a characteristic oscillation timescale and chaotic temporal fluctuations. Additional statistical and spectral analyses are provided in Fig.~\ref{Fig_3}(a).

\subsection{Estimation of Holevo capacities}
\label{subsec:holevo_estimation}

We outline the numerical procedure taking the quantum memory capacity $\chimem = S(\bar{\rho}_{t_{n+1}}) - \sum_{s_n} P(s_n) S(\rhomem)$ [Eq.~\eqref{eq:chi_memory}] as an example. The unconditional average state is $\bar{\rho}_{t_{n+1}} = \sum_{s_n} P(s_n) \rhomem$ [Eq.~\eqref{eq:consistency_check}], and the conditional memory state $\rhomem$ aggregates the reservoir configurations over all historical trajectories $S_{n-1}$ that lead to the present signal $s_n$: $\rhomem = \sum_{S_{n-1}} P\left(S_{n-1} | s_n\right) \rho_{t_{n+1}}\left(S_n\right)$ [Eq.~\eqref{eq:def_memory_state}]. 
Here, the state $\rho_{t_{n+1}}\left(S_n\right)$ evolves iteratively under the discrete thermalization map [Eq.~\eqref{eq:collisional_map}], expressed as $\rho_{t_{n+1}}(S_n) = \mathcal{E}_{t_n}(\rho_{t_n}(S_{n-1}))$

Evaluating these expressions requires the underlying probability distributions, such as $P(s_n)$ and $P(S_{n-1} | s_n)$. For a continuous chaotic driving signal like the MG sequence, these probabilities are governed by the continuous invariant measure of the chaotic attractor. According to ergodic theory, the natural probability measure of a chaotic system can be statistically reconstructed from a sufficiently large ensemble of independent trajectories \cite{Eckmann_1985, Kantz_2003}. Therefore, rather than tracking a single exceedingly long sequence, we simulate an ensemble of $5000$ independent time series to broadly sample the attractor space. 
To extract the probability densities and calculate the continuous entropy integral from this ensemble, we utilize a histogram estimation protocol. This approach is established as a robust method for estimating the entropy and mutual information of continuous distributions \cite{Moddemeijer_1989, Paninski_2003}. The continuous state space $[-1, 1]$ of the input signals is uniformly partitioned into $B = 50$ discrete bins. At each temporal step $t_{n+1}$, the density matrices $\rho_{t_{n+1}}\left(S_n\right)$ are assigned to a specific bin $b$ based strictly on the instantaneous scalar value of their corresponding driving signal $s_n$. 
This binning procedure directly resolves the required mathematical components. The classical probability $P(s_n)$ is approximated by the relative frequency $P_b$ of trajectories falling into bin $b$. Concurrently, averaging the density matrices within bin $b$ physically executes the conditional statistical summation described in Eq.~\eqref{eq:def_memory_state}, yielding the discrete conditional state $\rho_b \approx \rhomem$. The conditional entropy is thus approximated by a discrete sum:
\begin{equation}\label{eq:entropy_integral}
	\sum_{s_n} P(s_n) S(\rhomem) \approx \sum_{b=1}^{B} P_b S\left(\rho_b\right).
\end{equation}
Using $\bar{\rho}_{t_{n+1}} \approx \sum_{b=1}^{B} P_b \rho_b$, the memory capacity $\chimem$ is directly computed. The predictive capacity $\chipred$ is evaluated following an identical procedure, utilizing bins structured around the future target $s_{n+1}$.

This approach ensures both theoretical validity and computational reliability. Setting $B = 50$ provides sufficient resolution to capture the fine structural details of the continuous chaotic attractor while maintaining an average of $100$ independent samples per bin. This dense population effectively suppresses finite-sampling statistical fluctuations during the construction of $\rho_b$. We additionally verified that the numerical capacities rigorously converge with respect to the number of bins. Finally, to guarantee numerical stability and eliminate artificial singularities caused by zero eigenvalues during the logarithmic evaluation of the von Neumann entropy, a stringent eigenvalue truncation threshold of $10^{-12}$ is applied prior to calculating $S\left(\rho_b\right)$.

\bibliography{Refs}

\end{document}